\documentclass[10pt,a4paper]{article}
\usepackage{amsmath}
\usepackage{amsfonts}
\usepackage{amssymb}
\usepackage[left=1.5cm, right=1.5cm, top=2.00cm, bottom=2.00cm]{geometry}
\usepackage{natbib}
\usepackage{booktabs}
\usepackage{setspace}
\usepackage{subfigure}

\usepackage{graphicx}

\usepackage{fancyhdr}
\usepackage{rotating}
\usepackage{array}
\usepackage{color}
\usepackage{multirow}
\usepackage{float}
\usepackage{url}
\usepackage{subfigure}
\usepackage{hyperref}
\usepackage{abstract}
\usepackage{multicol}
\hypersetup{
unicode=false,          
pdftoolbar=true,        
pdfmenubar=true,        
pdffitwindow=false,     
pdftitle={Dayglow on Mars},    
pdfauthor={Sonal Kumar Jain},   
pdfsubject={Cameron band and UV doublet band},   
colorlinks=true,        
linkcolor=blue,         
citecolor=blue,        
filecolor=magenta,      
}

\newcommand{\car}{{$\mathrm{CO_2} $}}
\newcommand{\carp}{{$\mathrm{CO_2^+} $}}

\newcommand{\nt}{{$\mathrm{N_2}$}}
\newcommand{\dgr}{{$^\circ$}}
\newcommand{\coa}{CO(a$^3\Pi$)}
\newcommand{\carpb}{CO$_2^+(\mathrm{B}^2\Sigma_u^+)$}

\title{\bf Impact of solar EUV flux on CO Cameron band and CO$_2^+$ UV
doublet emissions in the dayglow of Mars}
\date{}
\author{Sonal Kumar Jain\thanks{sonaljain.spl@gmail.com} and
Anil Bhardwaj\thanks{anil\_bhardwaj@vssc.gov.in; bhardwaj\_spl@yahoo.com} \\ 
Space Physics Laboratory,\\ 
Vikram Sarabhai Space Centre,\\ 
Trivandrum, India - 695022\\
{\small Planetary and Space Science, 2011, doi:10.1016/j.pss.2011.08.010}}

\begin{document}

\maketitle

\begin{abstract}
This study is aimed at making a calculation about the impact of the two most
commonly used solar EUV flux models -- SOLAR2000 (S2K) of
\cite{Tobiska04} and EUVAC model of \cite{Richards94} -- on photoelectron
fluxes, volume emission rates, ion densities and CO Cameron and CO$_2^+$ UV doublet
band dayglow emissions on Mars in three
solar activity conditions: minimum, moderate, and maximum.
Calculated limb intensities profiles are compared with SPICAM/Mars Express and 
Mariner observations.
Analytical yield spectrum (AYS) approach has been used to calculate 
photoelectron fluxes in Martian upper atmosphere. Densities of prominent
ions and CO molecule in excited triplet a$^3\Pi$ state are calculated using
major ion-neutral reactions. Volume emission rates of CO Cameron and CO$_2^+$ UV
doublet bands have been calculated for dif{}ferent observations (Viking condition,
Mariner and Mars Express SPICAM observations) on Mars. For the low
solar activity condition, dayglow intensities calculated using the S2K model are
$\sim$40\% higher than those calculated using the EUVAC model. During high solar
activity, due to the higher EUV fluxes at wavelengths below 250~\AA\  in the EUVAC
model, intensities calculated using EUVAC model are slightly higher ($\sim$20\%)
than those calculated using  S2K model. Irrespective of the solar activity
condition, production of Cameron band due to photodissociative excitation of
CO$_2$ is around 50\% higher when  S2K model is used.
Altitude of peak limb brightness of CO Cameron and
CO$_2^+$ UV doublet band is found to be independent of solar EUV flux models.
Calculated limb intensities of CO  Cameron and CO$_2^+$ UV doublet bands are on an
average a factor of $\sim$2 and $\sim$1.5, respectively, higher than the SPICAM Mars
Express observation, while they are consistent with the Mariner observations.
\end{abstract}

\begin{multicols}{2}
\section{Introduction}\label{sec:introduction}

First  observation of CO Cameron and \carp\ UV doublet  emissions
in the Martian dayglow were made by the Mariner 6 and 7 flybys in 1969--1970
\citep{Barth71,Stewart72a}. These observations provided an opportunity to
study the Martian upper atmosphere in a greater detail.
The CO Cameron band ($a^3\Pi-X^1\Sigma^+$; 180 -- 260 nm) system arises due
to the transition from the excited triplet a$^3\Pi$ state, which is the lowest of
triplet states, to the ground state X$^1\Sigma^+$ of CO.
Doublet transition ($B^2\Sigma^+-X^2\Pi $) from excited \carp\ (B$ ^2\Sigma_u $)
to the ground state \carp\ (X$^2\Pi$) gives emission in ultraviolet wavelengths
at 288.3 and 288.6 nm.
Apart from these emissions,
Fox-Duf{}fenback-Berger band of \carp\ (A$ ^2\Pi_u $ -- X$ ^2\Pi_g $), fourth
positive band of CO, first negative band of CO$^+$ (B -- X), and several
atomic line emissions of carbon and oxygen atoms were also recorded by Mariner
6, 7, and 9 \citep{Barth71,Stewart72a,Stewart72b}. With the help of theoretical
calculations and laboratory measurements, \cite{Barth71} proposed
possible mechanisms for the dayglow emission observed in the Martian
atmosphere. Maximum intensity of CO Cameron band and UV doublet
observed by Mariner 6 and 7 were 600 kR at $ \sim $131 km and 35 kR at
148 km, respectively.

Emissions from Cameron band and \carp\ UV doublet bands were also observed
in 1971--1972 by Mariner 9, the first spacecraft to orbit Mars. 
\cite{Stewart72b} observed a reduction in the intensity of Cameron band by a factor
of 2.5 compared to Mariner 6 and 7 observations. They attributed this dif{}ference
to the reduction in the solar activity and better calibration of Mariner 9
instrument. The observed  maximum slant intensities of CO Cameron band were
between 200 and 300 kR and averaged topside scale height for the same band
was 17.5 km. \cite{Stewart72a} also observed a good correlation between
CO Cameron band intensity and solar F10.7 flux, which suggest that these
emissions are controlled by the incident solar photon flux.

Since the Mariner 6, 7, and 9 UV observations, SPICAM (SPectroscopy for 
the Investigation of the Characteristics of the Atmosphere
of Mars) on-board Mars Express  (MEX) is the first instrument dedicated
for the aeronomical studies of Mars. SPICAM  has broaden our understanding about
the Martian dayglow.  Emissions observed by SPICAM in UV dayglow are
H~Lyman-$\alpha$ emission at 121.6 nm, the atomic oxygen emissions at 130.4 and
297.2 nm, the Cameron band ($a^3\Pi - X^1\Sigma^+$)  and fourth positive band
($ A^1\Pi - X^1\Sigma^+ $) of CO, and ultraviolet  doublet band ($ B^2\Sigma^+-X^2\Pi $)
emissions of \carp\ \citep[cf.][]{Leblanc06,Chaufray08}. These  emission features
are similar to those observed by Mariner 6, 7, and 9 but with better sensitivity,
and spatial and temporal coverage. SPICAM has observed these dayglow emissions on Mars
throughout the Martian year and showed the ef{}fect of solar zenith angle (SZA),
seasonal variation, and Martian dust storms on the dayglow emissions
\citep{Leblanc06,Leblanc07,Shematovich08,Simon09,Cox10}. SPICAM also provided
first observation of \nt\ UV emissions in Martian dayglow \citep{Leblanc06, Leblanc07}.
\nt\  Vegard-Kaplan VK (0, 5) and (0, 6) band emissions at 260.4 nm and 276.2 nm,
respectively, have been observed; \nt\ VK (0, 7) emission at 293.7 nm has also been
reported, but it has large uncertainty \citep{Leblanc07}. The detailed model
of \nt\ dayglow emissions on Mars is presented elsewhere \citep{Jain11}.

Several theoretical studies have been made for the dayglow emissions on
Mars \citep{Fox79,Mantas79,Conway81,Shematovich08,Simon09}. First detailed
calculation of dayglow emission on Mars was carried out by \cite{Fox79}.
Calculated overhead intensities of CO Cameron and \carp\ UV doublet bands
were 49 kR and 12 kR, respectively, for the low solar activity condition
similar to Viking landing \citep{Fox79}. \cite{Simon09} used Trans-Mars model
to calculate limb intensities of Cameron and \carp\ UV doublet emissions for
low solar activity condition (similar to Viking landing) and compared them
with SPICAM-observations. Their calculated intensities are higher by $\sim$25\%
than the observation. \cite{Simon09} had to reduce the Viking \car\ density in
the model atmosphere by a factor of 3 to bring the altitude of peak emission
in agreement with the observation.

Seasonal ef{}fects on intensities of Cameron and UV doublet bands have been
observed by SPICAM \citep{Simon09,Cox10}. \cite{Cox10} have presented a
statistical analysis of Cameron band and UV doublet emissions, peak altitude
of emissions, and ratios between UV doublet and Cameron band. Averaged peak
emission brightness (altitude of peak emission) observed by \cite{Cox10} for CO Cameron
and UV doublet bands are 118 $ \pm $ 33 kR (121 $ \pm $ 6.5 km) and 21.6 $
\pm $ 7.2 kR (119.1 $ \pm $ 7.0 km), respectively. They also presented
observations for one particular season, solar longitude (Ls) = 90 to 180\dgr,
and compared observational data with model calculations based on Monte Carlo
code, which has been used also by \cite{Shematovich08} for the Martian dayglow
studies.

Modelling of CO Cameron and \carp\ UV doublet dayglow emissions requires
a sophisticated input solar EUV ($ \sim $50 to 1000 \AA) flux, which is a
fundamental  parameter to model physics, chemistry  and dynamics of the upper
atmosphere of planets. Since observations of solar EUV irradiance are not
frequent and generally not available simultaneously with the observation
for the upper atmospheric studies, the solar EUV flux model become important
for modelling of aeronomical quantities in planetary atmospheres. Generally,
solar EUV flux models are bin-averaged into numbers of wavelength bands and
important solar emission lines appropriate for the calculation of photoionization
and photoelectron impact production rates.
Characterisation of the solar EUV flux for use in aeronomical and ionospheric
studies were developed during the seventies based on the Atmospheric Explorer-E
(AE-E) data \citep{Hinteregger76,Hinteregger81,Torr85}. Two AE-E reference
spectra SC\#21REF and F79050N have been published by \cite{Torr85} at 37
wavelength bins for solar minimum and maximum conditions, respectively.
Later, based on the measured photoelectron flux, the short wavelength fluxes
were increased by \cite{Richards94}, and they incorporated modified F74113
solar EUV flux in their EUVAC model. Detailed discussion on the development of
solar EUV flux models is beyond the scope of this study; \cite{Lean90},
\cite{Richards94}, and \cite{Lean11} have provided reviews on solar EUV
flux models.

For a given solar activity there are significant dif{}ferences between
the EUV fluxes reproduced by dif{}ferent solar flux models. These models
are based on the dif{}ferent input parameters and proxies, e.g., solar 10.7 cm
radio flux (henceforth referred to as solar F10.7) is used as the
measure of solar activity, and used for parametrization of solar EUV flux models
\citep{Richards94,Tobiska90}. EUVAC model of  \cite{Richards94} is based on solar
F10.7 and its 81-day average and also on the F74113 solar flux (in the
EUVAC model, the F74113 flux below 250 \AA, and  below 150 \AA, are doubled and
tripled, respectively). SOLAR2000 model of \cite{Tobiska00} is based on measured
solar flux irradiance and various proxies and provides solar flux from X-rays to
infrared wavelengths, i.e., 1--1000000 \AA.

Dif{}ferent solar EUV flux models have been used
to study the  solar radiation interaction with upper atmosphere of Mars.
Presently, SOLAR2000 (S2K) model of \cite{Tobiska04} and EUVAC model of
\cite{Richards94} are commonly used solar EUV flux models in aeronomical
studies of Mars; {\it e.g.}, \cite{Simon09} and \cite{Huestis10} have used
EUVAC model, while \cite{Shematovich08} and \cite{Cox10} have used S2K model
for the dayglow calculations on Mars.

\cite{Fox96} have studied the ef{}fect of dif{}ferent solar EUV flux
models on calculated electron densities for low and high solar activity
conditions. They have used 85315 and 79050 solar fluxes of 
\cite{Tobiska91} and SC\#21REF and F79050N AE-E reference solar
spectra of Hinteregger \citep{Torr79} for low and high solar activity
conditions, respectively. \cite{Fox96} found that due to smaller fluxes
at short wavelength range (18--200 \AA) in Hinteregger spectra, lower 
peak in  electron density profile is significantly reduced (30--35\%)
compared to that calculated using solar fluxes of \cite{Tobiska91}.
\cite{Buonsanto95} calculated ionospheric electron density using
EUVAC model of \cite{Richards94} and EUV94X solar flux model of
\cite{Tobiska94}. They found that photoionization rate in F2 region 
calculated by using EUV94X model is larger than that calculated using
EUVAC model due to the large EUV fluxes in 300--1050~\AA\ wavelength range
in EUV94X solar flux model.
Recently, \cite{Jain11} have studied the ef{}fect of solar EUV flux models
on \nt\ VK band intensities in Martian dayglow and showed that EUV flux models
does af{}fect the \nt\ dayglow emissions. Similar conclusion have been drawn for
\nt\ dayglow emission on Venus \citep{Bhardwaj11b}

The aim of the present study is to calculate the impact of solar EUV flux models
on CO Cameron band and \carp\ UV doublet band intensities in Martian dayglow.
We have used 37 bin EUVAC model of \cite{Richards94} and S2K version 2.36 of
\cite{Tobiska04} as the solar EUV fluxes. In these models, bins
consist of band of 50 \AA\ width each and few prominent solar EUV lines, and are
suf{}ficient for the modelling of photoionization and photoelectron flux
calculations \citep{Richards94,Simon09}. Photoelectron flux, volume excitation rates
and overhead intensities are calculated using both the solar EUV flux models for low,
moderate, and high solar activity conditions. Line of sight intensities for Cameron
band and UV doublet emissions are calculated and compared with the latest
observations by SPICAM instrument.

\section{Model}\label{sec:model}
Model atmosphere consists of five gases (\car, N$_2$, CO, O, and O$_2$).
Model atmosphere for solar minimum condition is taken from the Mars Thermospheric
General Circulation Model (MTGCM) \citep{Bougher99,Bougher00,Bougher04}. Model
atmosphere for high solar activity is taken from \cite{Fox04}. Photoabsorption
and photoionization cross sections for gases considered in the present study
are taken from \cite{Schunk00}, and branching ratios for ionization in dif{}ferent
states are taken from \cite{Avakyan98}.

Production mechanisms for \coa\ are photon and electron impact dissociative excitation
of \car, electron dissociative recombination of \carp, and electron
impact excitation of CO. Since $X^1\Sigma^+ \rightarrow a^3\Pi$ is a forbidden
transition, resonance fluorescence of CO is not an effective excitation mechanism.
\coa\ is a metastable state; \cite{Lawrence72a} has measured its lifetime as 7.5
$\pm$ 1 ms, consistent with the measurement of \cite{Johnson72}.
Individual band lifetime can vary, e.g., lifetime of \coa\ $ \nu = 0 $ level
is around 3 ms  \citep{Gilijamse07,Jongma97}. Due to its long lifetime, cross
section for the production of Cameron bands due to electron impact dissociation of \car\
is dif{}ficult to measure in the laboratory. Overall, Cameron band  cross section
can have an uncertainty of a factor of 2 \citep{Bhardwaj09}. \cite{Ajello71b}
reported relative magnitudes of the cross section for the (0, 1) band at 215.8 nm.
\cite{Erdman83} estimated the total Cameron band emission cross section of 
$ 2.4 \times 10^{-16} $ cm$ ^{2} $ at 80 eV, whereas Ajello estimated a value of
$ 7.1 \times 10^{-17} $ cm$ ^{2} $ at 80 eV. \cite{Bhardwaj09} have fitted the
e-\car\ cross sections producing CO Cameron band based on the estimated value of
\cite{Erdman83}.
In our study cross section for Cameron band production due to electron impact on \car\
is taken from \cite{Bhardwaj09}. 
Cross section for photodissociation of \car\ producing \coa\ is taken from
\cite{Lawrence72a}.
To calculate the production rate of \coa\ due to dissociative
electron recombination process we have calculated the density of electron and
major ions  by including ion-neutral chemistry in the model. The 
coupled chemistry model is similar to that adopted in our other studies
\citep{Bhardwaj96,Bhardwaj99a, Haider05}. Rate coef{}ficients for ion-neutral
reactions are taken from \cite{Fox01}. Recently, \cite{Seiersen03} have
measured recombination rates for the e-\carp\ collision, with yield of 0.87 for the
channel producing CO of which \coa\ production yield is 0.29
\citep{Skrzypkowski98,Rosati03}. Ion and electron temperatures are taken
from \cite{Fox09}. Ion and electron densities are
calculated under steady state photochemical equilibrium.
To calculate the density of \coa\ we have also included various sources
of loss of \coa\ in our coupled chemistry model, which are given in
Table~\ref{tab:rate-cameron}.

Major production sources of \carp\ in $ B $ state are
photon and electron impact ionization of \car.
Cross section for the formation of \carpb\ state due to electron impact ionization
of \car\ is taken from \cite{Itikawa02}, and cross section due to photoionization of
\car\ is based on the branching ratio taken from \cite{Avakyan98}. For other gases
electron impact cross sections have been taken from \cite{Jackman77}.

\begin{figure}
\centering
\includegraphics[width=30pc]{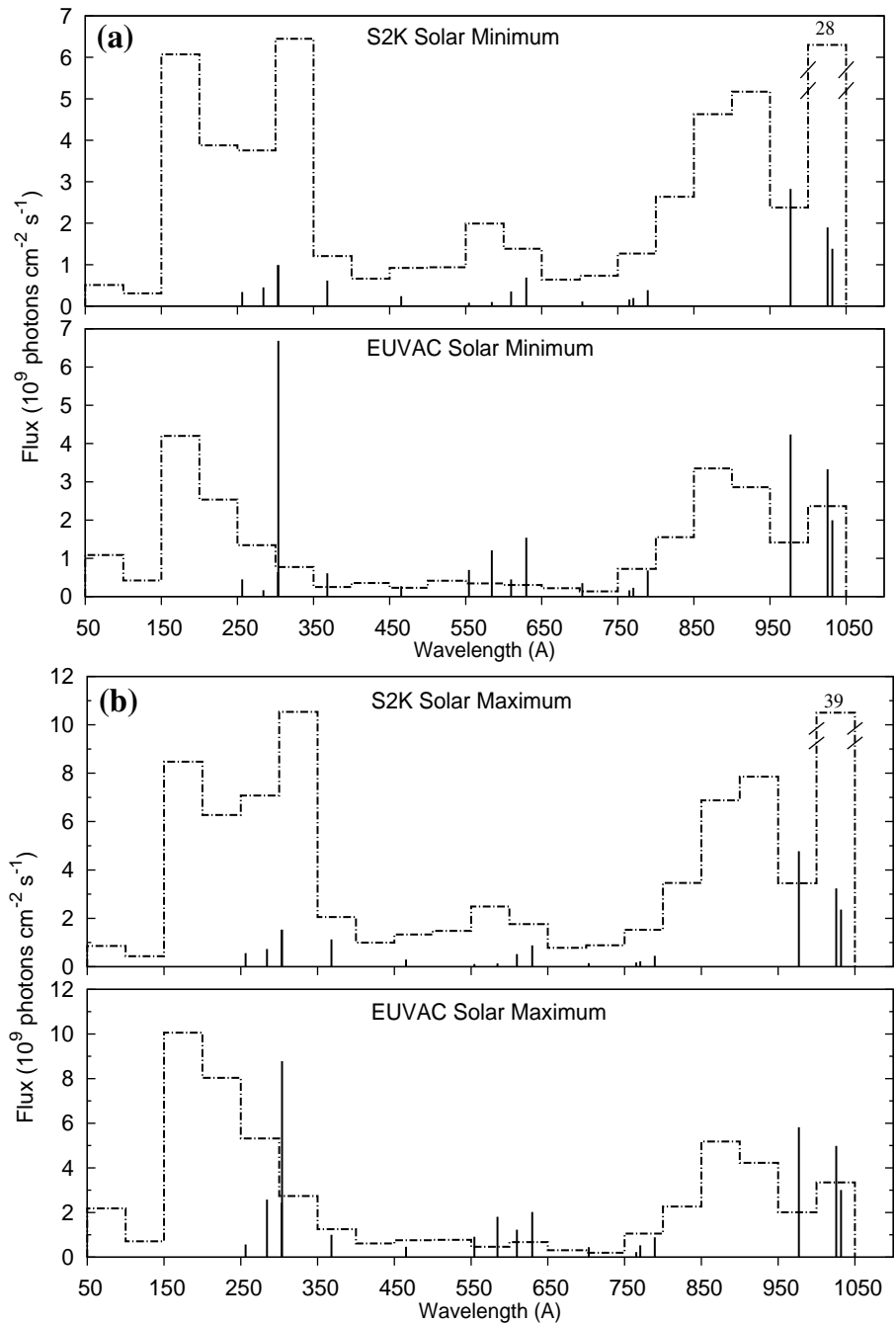}
\caption{Solar photon flux, in bands and at lines, as a function of wavelength in EUVAC and
S2K solar EUV flux models.
(a)  for the low solar activity condition on July 1976 (similar to Viking landing, F10.7 = 70),
and (b) for high solar activity condition on August 1969 (similar to Mariner 6 and 7
observations period, F10.7 = 186).}
\label{fig:solar-flx}
\end{figure}

Solar EUV flux has been taken at 1 AU (based on solar F10.7 at 1 AU as seen
from the Mars, taking the Sun-Earth-Mars angle into consideration) and then scaled
to the Sun-Mars distance. Fig.~\ref{fig:solar-flx} shows
the output of EUVAC and S2K solar EUV flux models for both solar minimum (20 July 1976)
and solar maximum (2 August 1969) conditions at 1 AU. There are substantial dif{}ferences in
the solar EUV fluxes of EUVAC and S2K models; moreover, these dif{}ferences are not similar
in solar minimum and maximum conditions. In both, solar minimum and maximum
conditions, solar flux in bands is higher in S2K than in EUVAC over the entire
range of wavelengths, except for bins below 250 \AA\ (150 \AA\ for solar minimum
condition), whereas solar flux at lines is higher in EUVAC model for entire 
wavelength range.
Overall higher solar fluxes above 250 \AA\ in S2K cause more photoionization.
Higher photon fluxes below 250 \AA\ (during solar maximum condition)
in EUVAC produce more higher energy electrons in the Martian atmosphere causing
secondary ionizations (cf. Fig.~\ref{fig:pef}) that can compensate for the higher
photoionization in S2K model. One big dif{}ference between solar EUV fluxes of S2K and
EUVAC  models is the solar flux at bin containing  1026 \AA\ H~Ly-$ \beta $
line, which in both  solar minimum and maximum conditions is around an
order of magnitude higher in S2K compare to EUVAC solar flux model
(cf. Fig.~\ref{fig:solar-flx}).
Flux at these wavelengths does not contribute to the photoionization, but are very
important for dissociative excitation processes. Cross section for the
photodissociation of \car\ producing Cameron band lies in longer (700--1080 \AA)
wavelength regime \citep{Lawrence72a}. Hence, yield of photodissociation excitation
of \car\ producing \coa\ state would be larger when S2K solar EUV flux model is used.

To calculate the photon degradation and generation of photoelectrons in the
atmosphere of Mars, we have used Lambert-Beer law. Solar zenith angle (SZA)
is  45\dgr\ unless mentioned otherwise in the text. Photoelectron energy
degradation and production rates of excitation of CO Cameron band and \carp\
UV doublet band in the Martian atmosphere are calculated using Analytical Yield
Spectrum (AYS) technique, which is based on the Monte Carlo model
\citep[cf.][]{Singhal91,Bhardwaj93,Bhardwaj99d,Bhardwaj99b,Bhardwaj09}. Details of
calculation of photoelectron production rates and photoelectron flux have given in
our earlier papers \citep{Bhardwaj90b,Bhardwaj03,Michael97,Jain11,Bhardwaj11b}.

Below 70 eV, photoelectron flux calculated using S2K is higher compared to that
calculated using EUVAC model for low solar activity condition (Fig.~\ref{fig:pef}).
Above 70 eV, photoelectron flux calculated using EUVAC model is higher than that
calculated using S2K model due to the larger solar EUV fluxes at shorter wavelength
($<$ 200~\AA) in EUVAC model (cf. Fig.~\ref{fig:solar-flx}). During solar maximum
condition photoelectron fluxes calculated by using EUVAC model is higher than that
calculated using S2K model (cf. Fig.~\ref{fig:pef}) due to higher solar
EUV flux at wavelength below 250 \AA\ in EUVAC model.
During solar minimum, except at peaks and energies above 70 eV, photoelectron flux
calculated using S2K model is around 1.4 times higher than that calculated using EUVAC model.
During solar maximum condition photoelectron flux calculated using EUVAC
model are higher than that calculated using S2K model.
Photoelectron fluxes calculated using both solar EUV flux models are similar 
in shape but peaks around 20--30 eV are more prominent in electron flux calculated
using EUVAC due to the higher solar EUV flux at lines (e.g., He II Lyman-$\alpha$ line
at 303.78 \AA) in EUVAC model (Fig.~\ref{fig:solar-flx}). The peaks in the 20--30 eV region
of the photoelectron flux arise due  to the ionization of \car\ in dif{}ferent ionization
states by solar EUV flux at 303.78 line \citep{Mantas79}. Our calculated photoelectron
fluxes are in good agreement with other model calculations \citep[cf.][]{Jain11}.

Volume excitation rate is calculated for important processes producing \coa\
and \carpb\ states using photoelectron flux as
\begin{equation}\label{eq:ver}
V_i(Z) = n(Z) \int _{E_{th}}^{E} \phi(Z, E) \sigma_i(E) dE,
\end{equation}
where $n(Z)$ is the density at altitude $ Z $, $ \sigma_i(E)$ is the electron 
impact cross section for $i^{th}$ process, for which threshold is
$E_{th}$, and $ \phi(Z, E) $ is the photoelectron flux.

\section{Results and discussion}
\label{sec:results}
\subsection{Low solar activity condition} \label{subsec:viking}
We run the model for low solar activity condition (similar to Viking landing), and
calculated results are compared with those of \cite{Fox79} by taking the similar
model atmosphere. Model atmosphere is based on density data derived from
Viking 1 \citep{Nier76}. The Sun-Mars distance D$_{\mathrm{S-M}}$ = 1.64 AU and
solar zenith angle is taken as 45\dgr.

Fig.~\ref{fig:ionden} shows the calculated densities of \carp\ and O$ _2^+ $
in the Martian upper atmosphere. The density of \carp\ around peak and above
calculated using S2K model is $\sim$30\% higher than that
calculated using EUVAC, which is due to higher production rate of \carp\ when
S2K model is used. Below 120 km, ion densities calculated by using EUVAC 
model are higher due to the higher photoelectron fluxes above 70 eV
(cf. Fig.~\ref{fig:pef}). There is a small discontinuity in the density of
O$ _2^+ $ ion around 180 km, which is due to the sudden change in the electron
temperature at 180 km \cite[see Fig.~2 of][]{Fox09}. Our calculated ion densities
are consistent with calculations of \cite{Fox04}.

Fig.~\ref{fig:ver-viking} (upper panel) shows the profiles of production
mechanisms of CO(a$^3\Pi$)  calculated using EUVAC and S2K solar EUV flux models.
Around the peak of \coa\ production, the major source is photoelectron
impact dissociation of \car, while at higher altitudes photodissociation
excitation of \car\ takes over. Dissociative recombination is about 13\%,
while photoelectron excitation of CO contribute about 3\% to the 
total Cameron band excitation at the peak. The shape of volume excitation rate (VER)
profiles and the altitude at the peak remain the same for all processes for
the two solar flux models. However, the magnitude of VERs calculated using S2K model
are about 40\% higher than those calculated using EUVAC model. Contribution of
electron impact dissociation of \car\ producing \coa\ is higher in our studies
than that in \cite{Fox79}. This is due to the higher value of \coa\ production cross
section in e-\car\ collision used in the present study (having the value of
$1\times10^{-16}$ cm$^{2}$ at 27 eV); \cite{Fox79} used the cross section
derived from \cite{Freund71} (having the value of $ 4\times 10^{-17} $ cm$^{2}$
at 27 eV). Due to larger photon flux at longer (700--1050 \AA) wavelengths
(region where photodissociation of \car\ becomes important) in S2K model compared
to EUVAC model (cf. Fig.~\ref{fig:solar-flx}), \coa\ production due to
photodissociative excitation is higher by $\sim$50\% when S2K model is used.
Production rates of CO Cameron band for different processes calculated at the
peak along with the peak altitude are given in Table~\ref{tab:ver}.

It is also clear from upper panel of Fig.~\ref{fig:ver-viking} that the altitude
where the photodissociation of \car\ takes over electron impact dissociation of
\car\ in  \coa\ formation is slightly higher when EUVAC model is used (167 km for
EUVAC and 160 km for S2K solar flux model).
In our model calculations, as well as in
the work  of \cite{Simon09}, photodissociation process becomes the major source at
higher altitudes ($>$ 160 km) and is a factor of 2 higher than the electron impact
dissociation of \car.

For the \carp\ UV doublet band, we have considered only photoionization and electron
impact ionization of \car\ producing \carp\ in the B$^3\Sigma^+_u$ state. Contribution
of solar fluorescent scattering is very small, less than 10\% \citep[cf.][]{Fox79}, and hence
it is not taken into account in the present study.  While calculating the emission from
\carp\ UV doublet, we have assumed 50\% crossover from $ B $ to $ A $ state \citep{Fox79}.
Fig.~\ref{fig:ver-viking} (bottom panel) shows the production rates for
\carp\ in $ B $ state. Production  of \carpb\ due to photoionization of \car\
is about a factor of 3--4 higher than due to photoelectron impact ionization.
Here also we find that production rates calculated using S2K are higher than
those calculated using EUVAC flux by about 50\%, but peak altitude remains the same
in both cases.
Production rates calculated at peak along with peak altitude for \carp\ UV doublet
band are given in Table~\ref{tab:ver}.

Volume emission rates are height-integrated to calculate overhead intensities,
which are presented in Table~\ref{tab:oi-cmp} for CO Cameron and UV doublet bands.
For Cameron band, contribution of e-\car\ process is maximum (64\%) followed
by photodissociation of \car\ (21\%). Contributions of dissociative recombination
and e-CO process are around 10\% and 3\%, respectively. For UV doublet band, the major
contribution is coming from photoionization of \car\ (80\%), the rest is due to
electron impact ionization of \car.

To compare the model output with observed emissions we have integrated volume 
emission rates along the line of sight. Limb intensity at each tangent point
is calculated as 
\begin{equation}\label{eq:los}
	I = 2 \int\limits_{0}^{\infty} \mathrm{V}(r)dr,
\end{equation}
where $r$ is abscissa along the horizontal line of sight, and V(r) is the volume
emission rate (in cm$^{-3}$ s$^{-1}$) at a particular emission point $ r $.
The factor of 2 multiplication comes due to symmetry along the line of sight 
with respect to the tangent point. While calculating limb intensity we assumed
that the emission rate is constant along local longitude/latitude. For
emissions considered in the present study the ef{}fect of absorption in the
atmosphere is found to be negligible.

Fig.~\ref{fig:int-viking} shows the calculated limb profiles of the CO Cameron
and \carp\ UV doublet bands along with the SPICAM-observed limb intensity taken from
\cite{Simon09}. Observed values are averaged over for the orbits close to Viking 1 condition
(Ls$\sim$100--140\dgr), low solar activity, and for SZA=45\dgr. Below 100 km there is
a sudden increase in the observed intensity of \carp\ UV doublet band, which, according
to \cite{Simon09}, is due to the very significant solar contamination below 100 km. Limb
intensities calculated using S2K flux are $\sim$40--50\% higher than those calculated
using EUVAC: clearly showing the ef{}fect of input solar EUV flux on the calculated
intensities. Magnitude of the calculated limb intensity of UV doublet band is in
agreement with the observation, but for CO Cameron band calculated intensity is higher
than the observed  profile. For both bands, the calculated intensity profile peaks at
higher ($\sim$5 km) altitude in comparison with the observation--indicating a denser neutral
atmosphere in our model.

The dashed curves in Fig.~\ref{fig:int-viking} show intensities calculated after reducing
the \car\ density by a factor of 1.5; a good agreement in the altitude of peak emission
is seen between calculated and observed limb profiles. Though the reduction in  \car\
density shifted the altitude of peak emission downwards, the magnitude of calculated
Cameron band  intensity is still larger than the intensity measured by SPICAM. 
As pointed out in Section~\ref{sec:model}, the e-\car\ cross sections producing
Cameron band are uncertain by a factor of $\sim$2. The calculated limb intensity
profile for reduced e-\car\ cross section by a factor of 2 is also shown in
Fig.~\ref{fig:int-viking}. Cameron band intensities obtained after reducing
the density and cross section are in relatively close agreement with the
observed values. In the model calculations of \cite{Simon09} also the Cameron band
intensity and its peak emission altitude were higher than the SPICAM
observed values. They have to reduced the density of \car\  by a factor of 3 and
e-\car\ Cameron production cross section by a factor of 2 to bring their calculated
intensity  profile in agreement with the SPICAM observation.

\subsection{SPICAM observations}\label{subsec:spicam-ana}
\cite{Leblanc06} have presented detailed analysis of SPICAM data during the period
October 2004 to March 2005, spanning the solar longitude (Ls) from 101\dgr\ to 171\dgr.
They divided the total data set in two periods of solar longitude: first,
Ls = 101\dgr\ to 130\dgr, and second, Ls = 139\dgr\ to 171\dgr\
\citep[cf. Table 2 of][]{Leblanc06}. \citeauthor{Leblanc06} found that the altitude of
peak emission for \carp\ UV doublet and CO Cameron bands is around 10 km higher
for Ls $>$ 138\dgr\ (122.5 km and 132.5 km for UV doublet and Cameron bands,
respectively) compared to Ls $<$ 130\dgr\ (112.5 km and 117.5 km, for the
same emissions). \cite{Leblanc06} could not provide the reason for the higher
altitude of peak emission for  Ls $>$ 130\dgr\ observations. Later, \cite{Forget09}
derived neutral densities in Martian upper atmosphere using the SPICAM
instrument in stellar occultation mode for the same observation period.
\cite{Forget09} found that their is a sudden increase in the \car\ density in the
Martian upper atmosphere for Ls $\sim$ 130\dgr--140\dgr,  which they attributed to a dust storm.
Dust storm can heat the lower atmosphere and thus increase the densities at higher
altitudes, which could explain the higher altitude for peak emission observed by
the SPICAM for Ls $ > $ 130\dgr\ observations. Increase in the altitude of peak
intensity of dayglow emissions clearly shows the ef{}fect of dust storms on Martian
dayglow emissions.

Comparisons of SPICAM observations with model calculated dayglow emissions
have been performed by \cite{Shematovich08}, \cite{Simon09}, and 
\cite{Cox10}. \cite{Simon09} have used one dimensional Trans-Mars model with
EUVAC solar flux model, whereas \cite{Shematovich08} and  \cite{Cox10} 
have used Monte Carlo model with S2K solar flux. In the present study,
we have taken both EUVAC and S2K models and have calculated the Cameron band
and UV doublet band emissions using the Analytical Yield Spectrum method;
the results are compared with the SPICAM observations.

\subsubsection{First Case (Ls $<$ 130\dgr)}
\label{subsubsec:firstcase}
To model the SPICAM observations for Ls $ < $ 130\dgr\ the model atmosphere is based
on MTGCM of \cite{Bougher99} \citep[taken from][]{Shematovich08}. Calculations 
are made for MEX orbit no. 983 (24 Oct. 2004) when D$_{\mathrm{S-M}}$ = 1.64 AU, 
F10.7 = 87.7 (F10.7A = 107.3).

Fig.~\ref{fig:ver-cam-shem} (upper panel) shows the volume excitation rate of \coa. The
total VER calculated using S2K flux is only slightly higher than that
obtained using EUVAC flux. However, the production rate due
to photodissociative excitation of \car\ is around 50\% higher when S2K model
is used. Another interesting feature is the dissociative recombination (DR)
process, whose contribution is $ \sim $18\% in the total intensity, roughly
equal to the photodissociative excitation process (DR contribution is even
higher than photodissociative excitation around production peak when EUVAC
model is used), and it is significantly higher than compared to the DR process
in Viking condition case (see Table~\ref{tab:oi-cmp}). This 
is due to the higher density of \carp\ ion compared to Viking condition
(see Fig.~\ref{fig:ionden}). \cite{Leblanc06} mentioned that
higher values of \carp\ can contribute up to 30\% to the Cameron band
production depending on the solar zenith angle. To account for
DR in Cameron band production, \cite{Shematovich08} and \cite{Cox10} 
have taken \carp\ and electron densities from \cite{Fox04} for low solar
activity condition. Since SPICAM observations are made during moderate
solar activity condition, the contribution of DR in Cameron band production
would be lower in their calculations.

Fig.~\ref{fig:ver-cam-shem} (bottom panel) shows the production rates of
\carp\ UV doublet band. Total rate calculated using both solar flux 
models is peaking at same altitude ($\sim$125 km), but total production rate
calculated using S2K model is higher (around 10\%) than that calculated
using EUVAC model.

Table~\ref{tab:oi-cmp} shows the overhead intensities of CO Cameron and 
\carp\ UV doublet bands calculated using both EUVAC and S2K solar flux models.
Contribution of dif{}ferent processes in Cameron band production is slightly
dif{}ferent than that in the low solar activity condition. Contribution from \car\ 
photodissociation is slightly reduced (17\%, 21\% when S2K model is used),
while dissociative recombination contribution is increased ($\sim$18\%).
Contribution of e-\car\ and e-CO processes remains almost same; 61 (64\%)
and 4 (4\%), respectively, when calculated using EUVAC (S2K) model. For UV doublet
band photoionization of \car\ remains the dominant process contributing around 80\%
to the total overhead intensity.

We have also calculated overhead intensities of major vibrational bands
of Cameron system, which have been observed in Martian dayglow, using Frank-Condon
factors from \cite{Halmann66} and branching ratio from \cite{Conway81}.
Table~\ref{tab:vib-oi} shows the calculated overhead intensities of major vibrational
bands of CO Cameron band system. Contribution of major vibrational bands to the total
overhead Cameron band intensity is around 10, 10, 16, and 8\% for (0, 0), (0, 1),
(1, 0), and (2, 0) bands, respectively.

Fig.~\ref{fig:int-min} shows the limb intensity profiles of Cameron 
and \carp\ UV doublet bands. SPICAM-observed intensities of Cameron
and UV doublet bands averaged over Ls = 100--130\dgr\ observations
\citep{Leblanc06} are also shown in the Fig.~\ref{fig:int-min}. Limb
intensities  of \carp\ UV doublet and Cameron bands calculated by using S2K
model are $\sim$6\% and $\sim$15\%, respectively, higher compared to
those obtained using EUVAC  model. Calculated intensities for both solar
flux models are higher than the SPICAM-observed values.
In analogy to the Viking case, we reduced the e-\car\ cross sections producing
Cameron band. The resulting intensity profile (also shown in
Fig.~\ref{fig:int-min}) is still higher than  the observation around emission peak.
Calculated intensity of \carp\ UV doublet band is also higher near the peak
emission than the observed intensity. Altitude of the calculated intensity
for both CO Cameron and UV doublet bands peaks $ \sim $2 to 3 km higher than
the observations, which is well within the observational uncertainties.
Line of sight intensity of dif{}ferent vibrational transitions of Cameron band at
the altitude of peak emission, which is $ \sim $120 km for first case, are shown
in Table~\ref{tab:vib-oi}.

Fig.~\ref{fig:ratio-min} shows the calculated intensity ratio of UV doublet to Cameron
band along with the observed ratio derived from  SPICAM observations \citep{Leblanc06}.
At lower altitudes calculated ratio is in agreement with observation 
($ \sim $0.18).
The ratio remains almost constant up to $\sim$120 km (where Cameron band and UV doublet
emission peaks), starts gradually decreasing with altitude and becomes almost constant
after 150 km. The observed ratio decreases almost monotonically from 100 km all
the way to 180 km. \cite{Leblanc06} have not found any dependence of SZA on the UV
doublet to Cameron band intensity ratios, though they have observed a weak dependence
of this intensity ratio on the solar activity. From the observed intensity ratio profile it
is clear that in upper atmosphere Cameron band intensity is increasing steadily compare
to UV doublet band, which indicates a dif{}ference in the source of production of
Cameron band and UV doublet band. That source could be the dissociative recombination
process which is sensitive to the density of \carp\ ion (as shown in the 
Fig.~\ref{fig:ionden}). Loss of \carp\ ions at higher altitudes ($>$ 200) can reduce
the intensity of UV doublet and hence decreases the intensity ratio value.

\subsubsection{Second Case (Ls $>$ 130\dgr)}
\label{subsubsec:secondcase}

As discussed earlier, due to dust storm during SPICAM observations for Ls greater 
than 130\dgr, atmospheric densities were higher resulting in altitude of peak
emission shifting to higher altitudes ($\sim$132.5 km for Cameron band emission).
For Mariner 6 and 7 observations the intensity of CO Cameron band peaked at altitude
of $\sim$133 km. Mariner observations were carried out during solar maximum condition
(F10.7 $\sim$ 180), whereas SPICAM observations are made during moderate solar
activity condition. To model dayglow emissions for Ls $ > $ 130\dgr, we
have made calculation  for MEX orbit 1426 (26 Feb. 2005), taking model atmosphere from
\cite{Fox04} for high solar activity condition. Sun-Mars distance is 1.5 AU,
F10.7 = 98 (F10.7A = 97). The EUV flux at 1 AU calculated using EUVAC model
remains the same for first (Ls $<$ 130\dgr) and second (Ls $>$130\dgr) cases.
This is because in the EUVAC model average of F10.7 and F10.7A (81-day
average) is used to scale the solar flux, and on both days average of F10.7 and F10.7A
does not change (it is 97.5 on both days), although the F10.7 flux increased by 10 unit.
S2K model does not depend on the F10.7 alone, but on other proxies also 
\citep{Tobiska04}, hence flux calculated using S2K model is dif{}ferent on the two days.

Fig.~\ref{fig:ver-cam-138} (upper panel) shows the VER of \coa\
calculated using S2K and EUVAC models. Total VER calculated using S2K is about 17\%
higher than that calculated using EUVAC model. Major dif{}ferences are in the \coa\
production due to dissociative excitation of \car\ by photon and electron impact,
which are more than 50\% and 10\% higher, respectively, when S2K model is used.
Total production rate of Cameron band (calculated using EUVAC model) maximises
at an altitude of 134 km with a value of about 3328 cm$^{-3}$ s$^{-1}$, which is around
10 km higher than that in the first case (Ls $<$ 130\dgr). Although production rate (3528
cm$^{-3}$ s$^{-1}$) at peak altitude is higher in the first case, but at higher
altitudes rate increases faster in the second (Ls $>$ 130\dgr) case, e.g., at 200 km, Cameron
band production rate is 61 cm$^{-3}$ s$^{-1}$ in second case, whereas in first case
it is only 3 cm$^{-3}$ s$^{-1}$. In both, first and second cases, for EUVAC model,
the altitude where photodissociation of \car\ takes over electron impact dissociation
of \car\ is around 30\% higher than that for S2K model.

Bottom panel of Fig.~\ref{fig:ver-cam-138} shows production rates of \carp\ UV
doublet band. Total excitation rate calculated using the S2K model is about 12\%
higher than  that calculated using EUVAC model. Similar to the Cameron band,
\carp\ UV doublet production rate at peak is lower than that in the first case,
but at higher altitudes UV doublet production rate becomes higher in the second case.
Table~\ref{tab:oi-cmp} shows the overhead intensities of Cameron and \carp\ UV 
doublet bands. Contribution of photodissociation of \car, e-\car, DR, and e-CO
processes to the total Cameron band production is 16 (22\%), 62 (57\%), 11 (11\%),
and 9 (8\%), respectively, when EUVAC (S2K) model is used.

Fig.~\ref{fig:int-max} shows the calculated limb intensity of UV doublet and 
Cameron band along with SPICAM-observed intensities of Cameron band and \carp\ doublet
band for MEX orbit 1426 on 26 Feb. 2005 \citep{Shematovich08}.
Intensities calculated using S2K model are higher by $ \sim $12--18\%
than those calculated using EUVAC model. Altitude of peak emission of calculated
and observed intensity profiles is in good agreement with each other
(e.g., $\sim$128 km for Cameron band) within the uncertainties of observations
and model calculations. However, intensities calculated using
both solar flux models are higher than the observations.
Calculated intensities of CO Cameron band after reducing the e-\car\ cross section by
a factor of 2 are also shown in Fig.~\ref{fig:int-max}, which is in good 
agreement with the observed values.
However, the calculated intensity of UV doublet  band is 20--30\% higher than the
SPICAM-observed values. Uncertainties in photoionization cross sections for the production
of \carpb\ can also be one of reasons for the dif{}ferences between observed and
calculated intensity of \carpb\ UV doublet emission.

In our calculation (using EUVAC model and reduced e-\car\ cross section for
Cameron band production), peak brightness of Cameron band
is  249 (255) kR with peak located at 120 (128) km in first (second) case. For \carp\ UV 
doublet band, peak brightness is 31 (33) kR with peak located at 119 (127) km in first
(second) case.
Altitude of peak brightness remains the same when S2K model is used but intensities of Cameron
and \carp\ UV doublet bands for first (second) case are about a factor of 1.1 (1.18) and 1.07
(1.11), respectively, higher than that calculated using EUVAC model.

\subsection{Solar Maximum (Mariner Observations)}
\label{subsec:mariner}
First dayglow measurements on Mars were carried out by  Mariner series of spacecraft.
Mariner 6 and 7  observations were taken during the solar maximum conditions (July-August,
1969; F10.7 = 186 at 1 AU), and are the only Martian dayglow observations so far taken
during the solar maximum condition. Mars was also at perihelion (distance between Sun
and Mars was around 1.42 AU) during Mariner observations. We run our model for the
condition similar to the Mariner observation. Model atmosphere for solar maximum
condition is taken from \cite{Fox04}.

Fig.~\ref{fig:solar-flx} (bottom panel) shows the solar EUV flux during Mariner
observations (2 Aug. 1969, F10.7 = 180) from EUVAC and S2K models. As discussed in
Section~\ref{sec:model}, overall the solar EUV flux is higher in S2K model, except at
wavelengths below 250 \AA\ where EUVAC flux is larger. Fig.~\ref{fig:pef} (bottom panel)
shows photoelectron fluxes  calculated  at dif{}ferent heights in solar maximum
condition using both solar flux models. Unlike in the solar minimum condition, the
ratio of photoelectron flux  calculated using S2K to that of EUVAC is less than 1
at most of the energies. At electron energies below 30 eV, fluxes calculated using
S2K and EUVAC models are almost equal at altitudes of 130 and 160 km. Above 30 eV,
photoelectron fluxes calculated using EUVAC model are higher than those calculated
using S2K model. The higher photon fluxes in EUVAC model at shorter wavelengths
below 250 \AA\ [cf. Fig.~\ref{fig:solar-flx}] produce higher energy photoelectrons
that can produce more secondary electron through ionization, and hence compensates
for the higher solar EUV flux above 250~\AA\ in the S2K model. Table~\ref{tab:oi-cmp}
shows the calculated overhead intensities of Cameron band and UV doublet band using
EUVAC and S2K models. Cameron band production due to electron impact dissociation of
\car\ is higher when EUVAC model is used, which is due to the higher photoelectron
fluxes. Contribution of photodissociation excitation calculated using S2K model
is higher, due to higher EUV flux at longer wavelengths (specially flux in the
1000--1050 \AA\ bin).

Fig.~\ref{fig:ver-cam-max} (upper panel) shows the VER of \coa\  for higher
solar activity condition, calculated using EUVAC and S2K solar
flux models. Due to the higher photoelectron flux, \coa\ production due to 
e-\car, e-CO, and dissociative recombination are higher when 
EUVAC model is used. Photodissociative excitation of \car\ producing
Cameron band for S2K model is still higher by $\sim$50\%
than for EUVAC model, which is due to the higher EUV fluxes at longer
wavelengths in the S2K model. Similar to that in the previous cases,
the cross over point between photodissociation and electron impact dissociation
of \car\ forming \coa\ occurs at higher altitude when EUVAC model is used.
Bottom panel of Fig.~\ref{fig:ver-cam-max}
shows the production rates of \carp\ UV doublet band. Here also 
calculated values using EUVAC model is slightly higher than that calculated
using S2K model.

During solar minimum condition total volume production rate of Cameron and UV doublet bands
calculated using S2K model is higher than
that calculated using EUVAC model, whereas in solar maximum it is vice-versa.
Except photodissociation excitation process producing Cameron band, production
rates due to other processes calculated by using EUVAC model are higher than
that calculated by using S2K model. In both, solar minimum and maximum conditions,
Cameron band production due to photodissociative excitation is about 50\% higher,
when S2K model is used.

Fig.~\ref{fig:int-mariner} shows model intensities  of Cameron band  and
\carp(B -- X) ultraviolet doublet band calculated using both EUVAC and S2K
models at SZA = 45\dgr\ along with  intensities observed by Mariner 6
and 7. Limb intensities  measured by Mariner 6 and 7 on Mars are at SZA = 27\dgr\
and 0\dgr, and at SZA = 44\dgr\ and 0\dgr, respectively. Calculated limb
intensities using EUVAC model at SZA = 0\dgr\ are also shown in the
Fig.~\ref{fig:int-mariner}. Limb intensities calculated
using EUVAC model are only slightly higher than those calculated using S2K model.
There were no observations at emission peak for both Cameron band and
\carp(B -- X) band. Below the emission peak, there are few observations  for
Cameron band, but ultraviolet doublet observations were not available.
Solar zenith angle ef{}fect is clearly visible at altitudes below 150 km, where
intensity is larger and emission peak shift deeper in the atmosphere for lower
SZA. Calculated intensities of Cameron and UV doublet bands are lower than
the observed values.
Unlike previous cases, calculated intensities of Cameron and UV doublet band
emissions at the altitude of peak emission are slightly lower than the observation.
\cite{Stewart72b} has pointed out that  due to calibration problem in Mariner 6
and 7 instrument the observed values can be higher.
As in the previous cases a reduction in e-\car\ cross section is required to
get an agreement between observed and calculated intensity. Calculated intensity
of Cameron band after reducing the e-\car\ cross section by a factor of 2 is also
shown in Figure~\ref{fig:int-mariner} for SZA=0\dgr. 

\subsection{\coa\ density}
\label{subsec:coa-den}
Density of \coa\ is calculated under photochemical equilibrium condition.
Radiative decay is the dominant loss process of \coa,
the contribution from  other processes are negligible \citep[cf.][]{Bhardwaj11a}.
Fig.~\ref{fig:ionden} shows the  density of \coa\ calculated using EUVAC and S2K EUV
flux models for low solar activity condition. The calculated column density of \coa\
molecule is $ 4.6 \times 10^7 $  ($ 6.5 \times 10^7 $) cm$ ^{-2} $ for the solar
minimum condition using EUVAC (S2K) solar EUV flux model. Except in the solar maximum 
condition, density of \coa\ molecule calculated using S2K model is higher
than that calculated using EUVAC model. During solar maximum condition, \coa\ 
density calculated using EUVAC model is slightly higher at peak (around 5\%), but at
altitudes above 140 km, density calculated using S2K model becomes higher ($\sim$10\%
at 200 km).

The shape of the density of \coa\ is similar to that of its production rate 
(cf. Fig.~\ref{fig:ver-viking})
since the main loss mechanism of \coa\ is radiative decay whose value is independent
of altitude. Hence, the
density of \coa\ in the Martian atmosphere can be represented by
\begin{equation}
[\mathrm{CO(a^3\Pi)}] = \frac{\mathrm{[CO_2]\,(K_1 + K_2) + [CO_2^+]\,[n_e]\, K_3}}{\mathrm{K_4}}
\label{eq:coa-den}
\end{equation}
where $K_1$, $K_2$, $K_3$, and $K_4$ are as described in Table~\ref{tab:rate-cameron}. $K_1$
and $K_2$ are photodissociation rate and  electron impact dissociation rate of \car, respectively,
$K_3$ is dissociative recombination, $K_4$ is radiative decay loss, and $n_e$ is the
electron density. The values of $K_1$
(photodissociation rate) in units of s$^{-1}$ at the top of atmosphere in case of EUVAC
(S2K) model are $7.5 \times 10^{-8}$ ($1.1 \times 10^{-7}$), $8.7 \times 10^{-8}$ ($1.3 
\times 10^{-7}$), $1.03 \times 10^{-7}$ ($1.6 \times 10^{-7}$), and $1.5 \times 10^{-7}$
($2.25 \times 10^{-7}$) in the solar minimum, first case, second case, and solar maximum,
respectively.

\section{Summary and Conclusion}
Present study deals with the model calculations of CO Cameron band and \carp\
doublet ultraviolet emissions in Martian dayglow and the impact of solar EUV
flux on the calculated intensities.
Photoelectrons generated due to photoionization in the Martian atmosphere
have been degraded in the atmosphere using Monte Carlo model--based Analytical
Yield Spectrum technique. Emission rates of Cameron and \carp\ UV doublet bands
due to photon and electron
impact on \car\ have been calculated using EUVAC and S2K solar EUV
flux models. Densities of prominent ions and \coa\ in Martian upper atmosphere
are calculated under steady state photochemical equilibrium condition.
Production rates of Cameron and \carp\ UV doublet bands are height-integrated
to calculate overhead intensity  and along the line of sight to obtain limb
intensities. Limb intensities are compared with the SPICAM/Mars Express and
UV spectrometer/Mariner observed intensities.

Due to higher EUV fluxes at longer (700--1050~\AA) wavelengths in the S2K model, the
contribution of photodissociation of \car\ in producing Cameron band is about 50\%
 higher in low as well as in high solar activity conditions. Variations in EUV fluxes
at longer wavelengths from solar minimum to solar maximum are less prominent in the
EUVAC solar EUV flux model compared to the S2K model.

For low solar activity condition, limb intensities of Cameron and \carp\ UV
doublet bands around peak brightness calculated using S2K model are $\sim$30--40\%
higher than those
calculated using EUVAC model. Comparison of calculated intensities has been
made with the SPICAM-observed values for condition similar to the Viking.
Intensities calculated using  both S2K and EUVAC models are higher than the observed
values. Calculated altitude of emission peak of CO Cameron and \carp\ UV doublet
bands is also higher by $\sim$5 km than the observed value. A reduction in the
e-\car\ cross section forming Cameron band by a factor of 2 and the density of
\car\ in model atmosphere by a factor of 1.5 brings the calculated intensity
(using EUVAC model) of Cameron band in close agreement with the SPICAM
observation.

While modelling the recent observations made by SPICAM on-board Mars Express, we 
have taken two set of conditions with dif{}ferent model atmospheres and solar
longitudes. In the first case, Ls $ < $ 130\dgr\ (Ls = 100-130\dgr), atmosphere
is taken from Mars Thermospheric General Circulation model 
\citep{Bougher99,Bougher00,Bougher04} and calculations are made for the day 24 October
2004 with moderated solar activity flux (F10.7 = 88). Total intensities of CO Cameron
and \carp\ UV doublet bands calculated using S2K model are around $\sim$6--15\% higher
than those calculated using EUVAC model. Contribution of \car\ photodissociative
excitation in Cameron band production is 50\% higher when S2K model is used.
Dissociative recombination of \carp\ is an important source of Cameron band 
in this case (cf. Fig.~\ref{fig:ver-cam-shem}) due to higher densities
of \carp\ ion (Fig.~\ref{fig:ionden}) compared to those calculated for low solar
activity condition (Viking condition).

Calculated intensities of Cameron and UV doublet bands have been
compared with the SPICAM-observed limb intensities. Intensities calculated
using S2K and EUVAC solar flux models are  higher than the observed values 
by a factor of 1.7 to 2 for Cameron  and a factor of 1.4 for UV doublet bands
(see Fig.~\ref{fig:int-min}).
We found that altitude of peak emission of both Cameron and UV doublet bands
are 2 to 3 km higher than observed profiles. This discrepancy in observed and
calculated intensities and altitudes could be due to the fact that observed values
are averaged over several days of observations while calculation are carried
out for a particular day.

Due to the dust storm during Ls $>$ 130\dgr, observed emission peak is around 10 km
higher for both Cameron and UV doublet compared with the SPICAM-observed
values for MEX orbit 1426 on 26 Feb. 2005 \citep{Shematovich08}.
To model the emission during Ls $ > $ 130\dgr, we have taken
atmospheric model for solar maximum condition. Intensities calculated using the 
S2K solar flux model are $\sim$8--18\% higher than those calculated using the EUVAC
model. These  calculated intensities are higher than the observed-averaged values
by a factor of $ \sim $2 for Cameron band and $ \sim $50\% for UV doublet band.
The calculated intensity of Cameron band (after reducing the e-\car\ cross sections
by a factor of 2) is in agreement with the observed values (Fig.~\ref{fig:int-max}).

In all three conditions discussed above, i.e., low solar activity (Viking), and first
(Ls $<$ 130\dgr) and second (Ls $>$ 130\dgr) cases, calculated
intensities of both Cameron and UV doublet bands are higher than observations.
On an average, model values of \cite{Cox10} for Cameron and \carp\ UV doublet bands
are around a factor of 1.74 and 1.41, respectively, higher than the SPICAM
observations. \cite{Simon09} also found that their calculated intensities of
Cameron and UV doublet bands are around 25\% higher than the SPICAM-observed
values. This shows that these discrepancies in the model and observed values are
due the uncertainties in the input physical parameters in the model. Uncertainties in
cross sections, namely, e-\car\ cross section producing Cameron band and
photoionization of \car\ forming UV doublet band can be one of the causes of 
discrepancies in the model and observations.

\begin{figure}
\centering
\includegraphics[width=30pc]{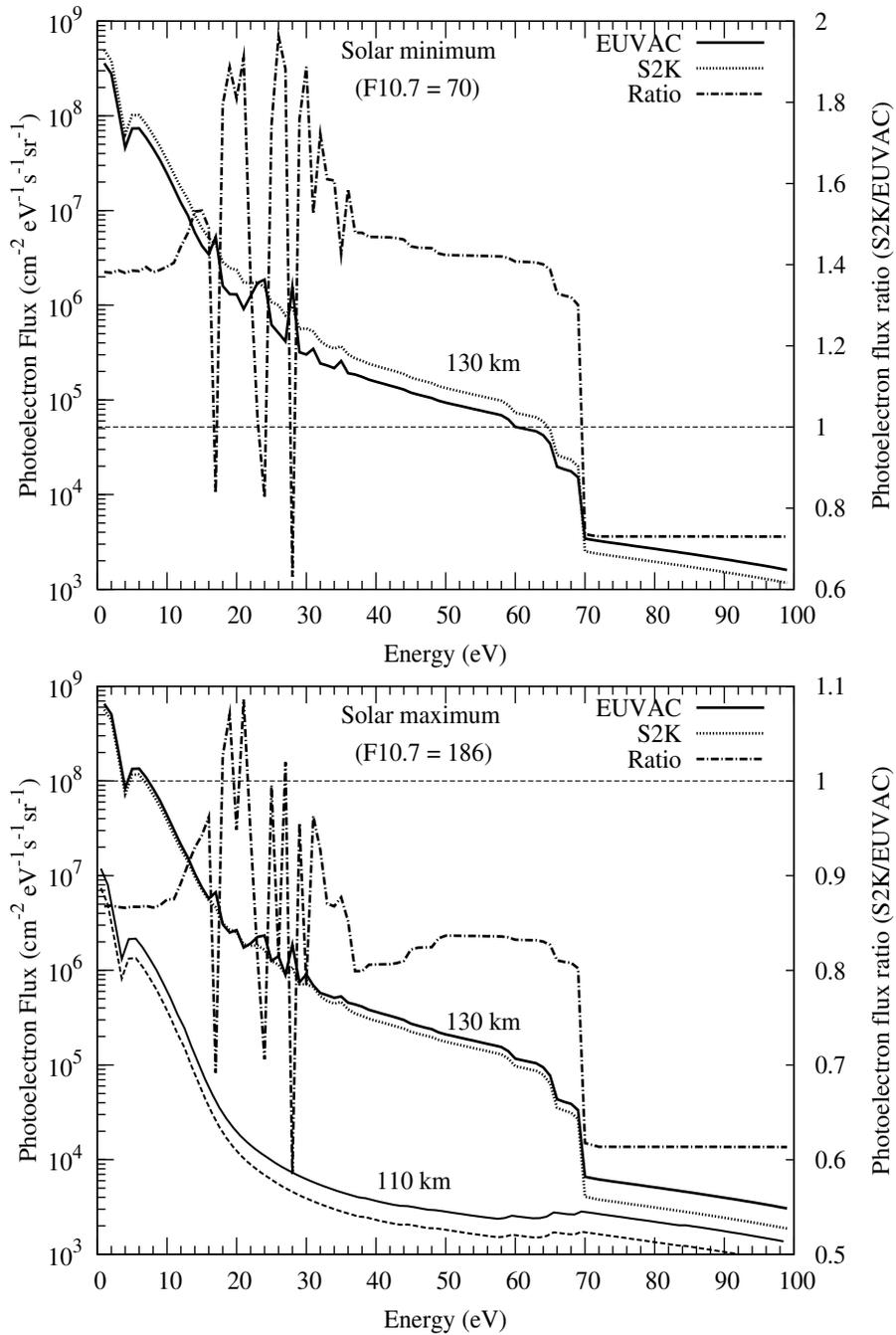}
\caption{Calculated photoelectron fluxes for low (upper panel) and high (lower panel)
solar activity conditions at SZA = 45\dgr. The ratio of the photoelectron flux at 130 km
calculated using the two solar flux models is also shown with magnitude on right side Y-axis.
Thin dotted horizontal line depicts the S2K/EUVAC ratio = 1.}
\label{fig:pef}
\end{figure}

\begin{figure}
\centering
\includegraphics[width=36pc]{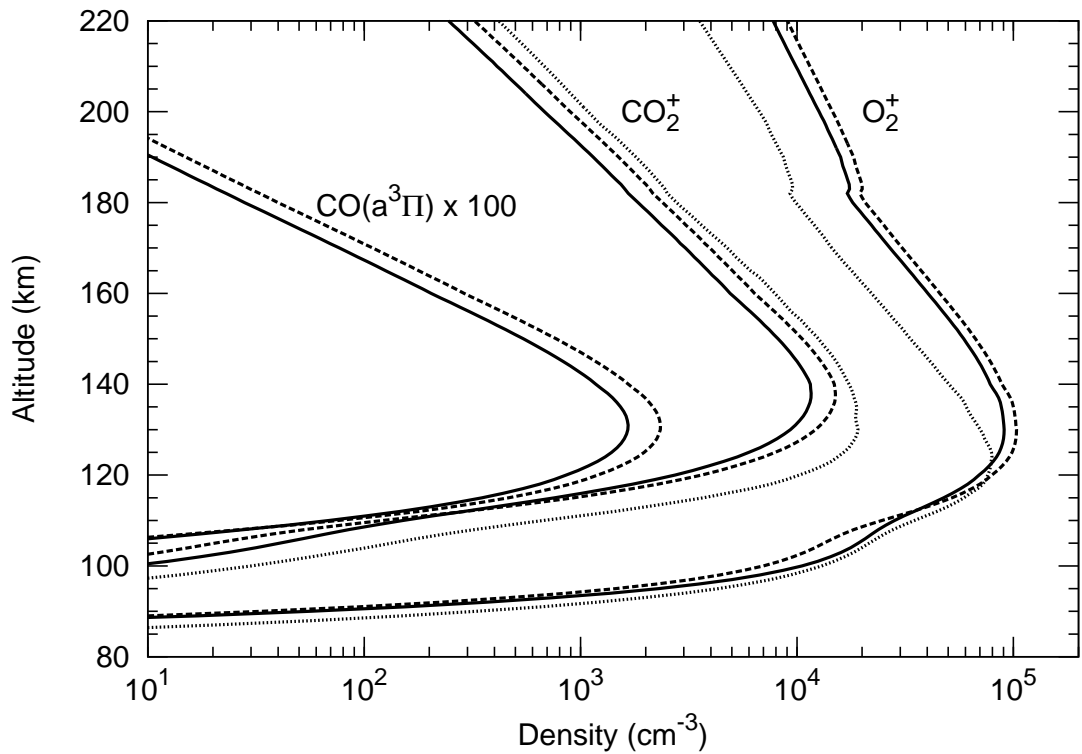}
\caption{Densities of \carp\ and O$ _2^+ $ ions and \coa\ molecule for solar minimum condition
calculated using EUVAC (solid curve) and S2K (dashed curve) solar EUV flux models.
Density of \coa\ molecule is plotted after multiplying by a factor of 100.
Dotted curves show the densities of \carp\ and  O$ _2^+ $ ions for first case
(Ls $ < $ 130\dgr) using EUVAC solar flux.} 
\label{fig:ionden}
\end{figure}

\begin{figure}
\centering
\includegraphics[width=30pc]{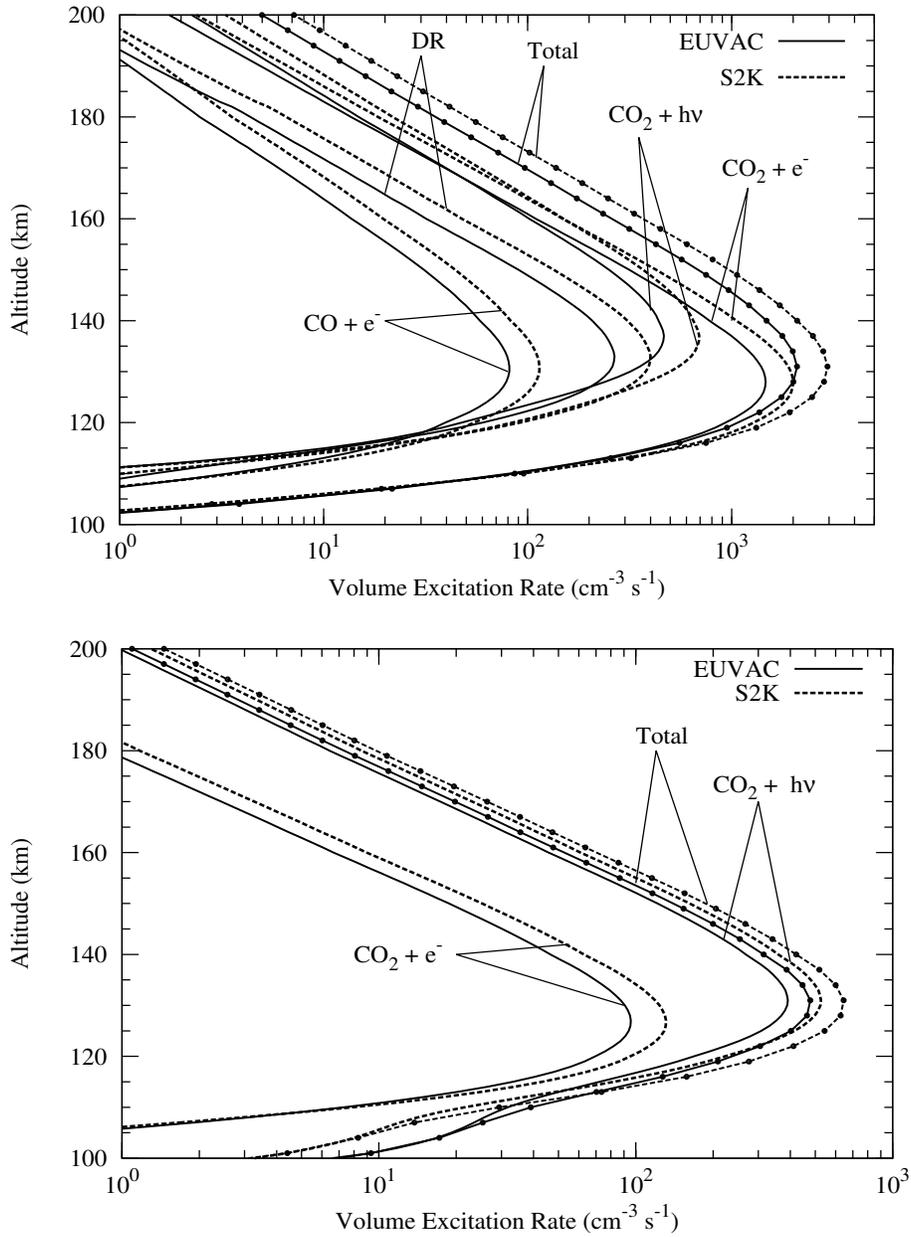}
\caption{Calculated production rates of the \coa\ (upper panel)  and \carpb\ (bottom panel)
for low solar activity condition (Ls$\sim$100--140\dgr). DR stands for
dissociative recombination.}
\label{fig:ver-viking}
\end{figure}

\begin{figure}
\centering
\includegraphics[width=36pc]{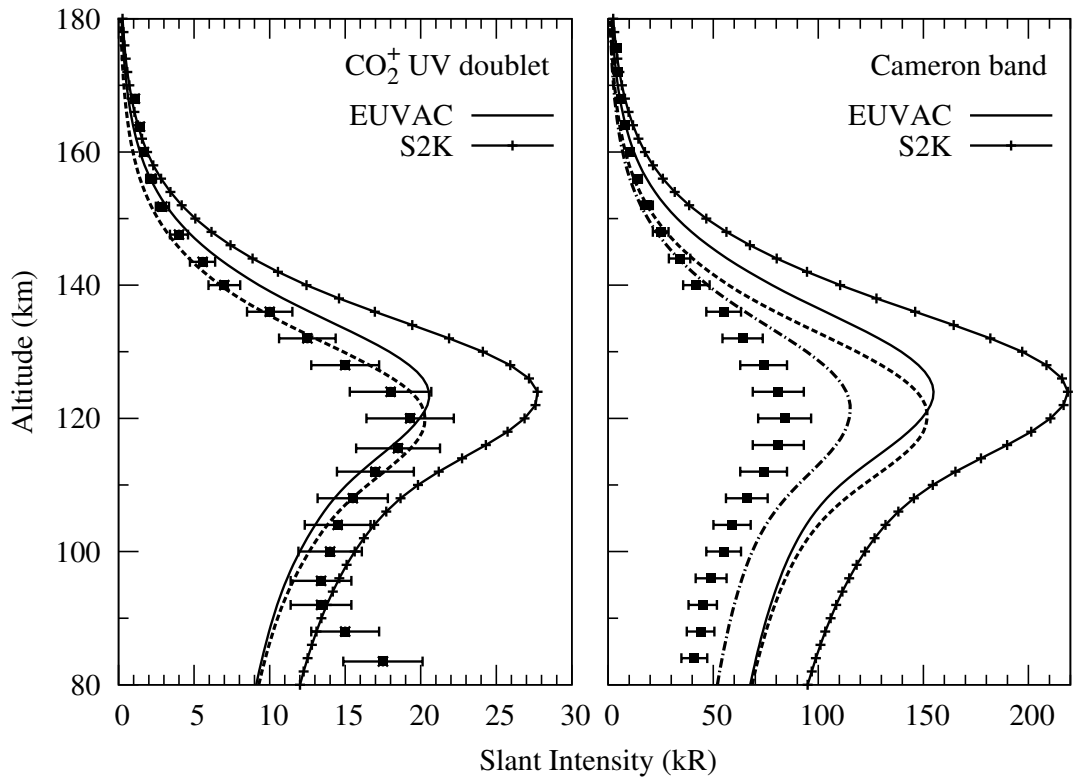}
\caption{Calculated limb profiles of \carp\ UV doublet bands (left panel) and
CO Cameron (right panel) for low solar activity condition. Solid squares with error
bars represents the SPICAM-observed values taken from \cite{Simon09}. Dashed curves
show the calculated intensity (using EUVAC model) after reducing the density of \car\
by a factor of 1.5. Dash-dotted curve shows the calculated intensity (using EUVAC) of
Cameron band with reduced density (by a factor of 1.5) and reduced (by a factor of 2)
e-\car\ Cameron band production cross section.} 
\label{fig:int-viking}
\end{figure}

\begin{figure}
\centering
\includegraphics[width=30pc]{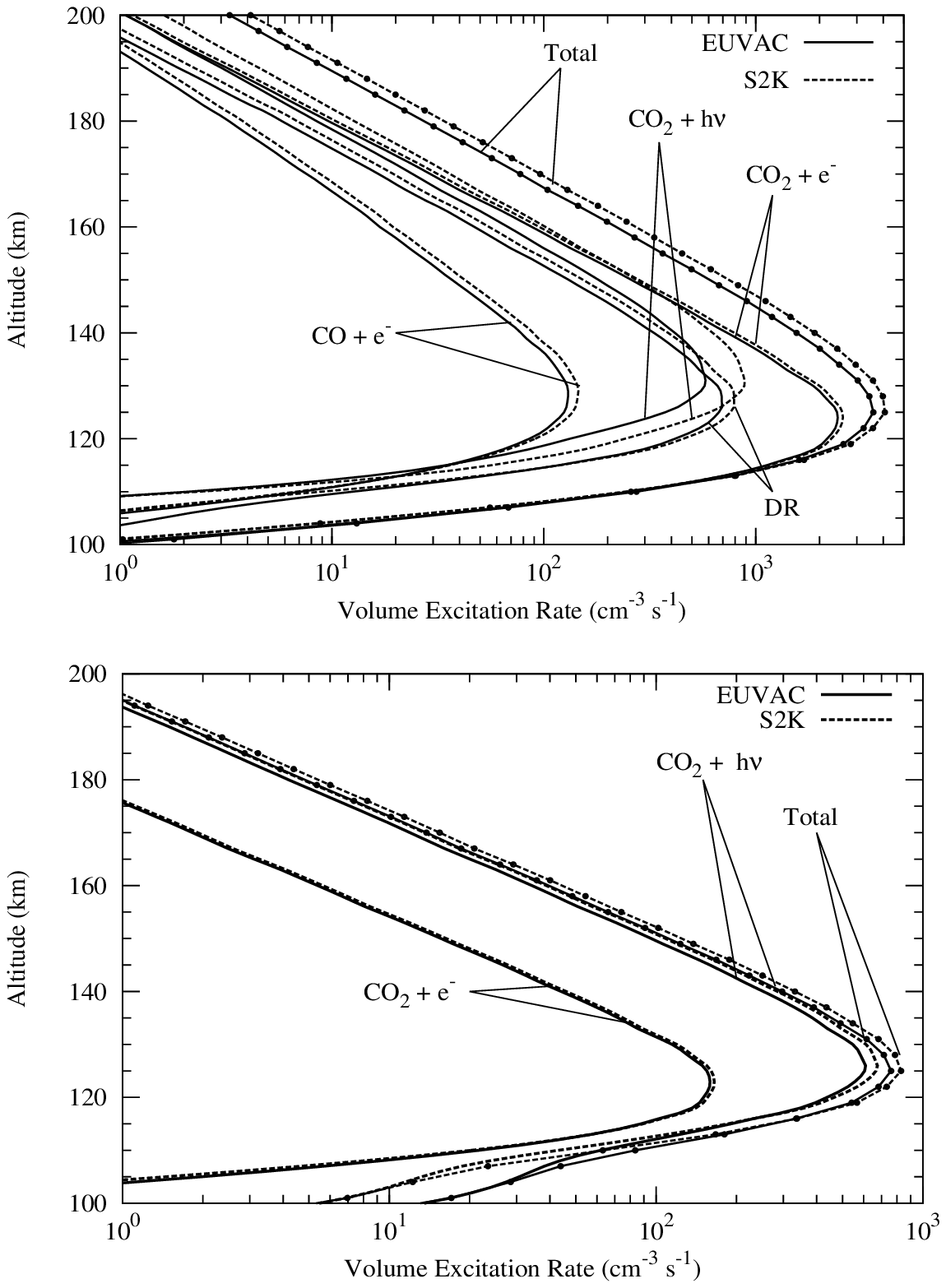}
\caption{Calculated production rates of the \coa\ (upper panel)  and \carpb\ (bottom panel)
for solar longitude Ls $<$ 130\dgr. DR stands for dissociative recombination.}
\label{fig:ver-cam-shem}
\end{figure}

\begin{figure}
\centering
\includegraphics[width=36pc]{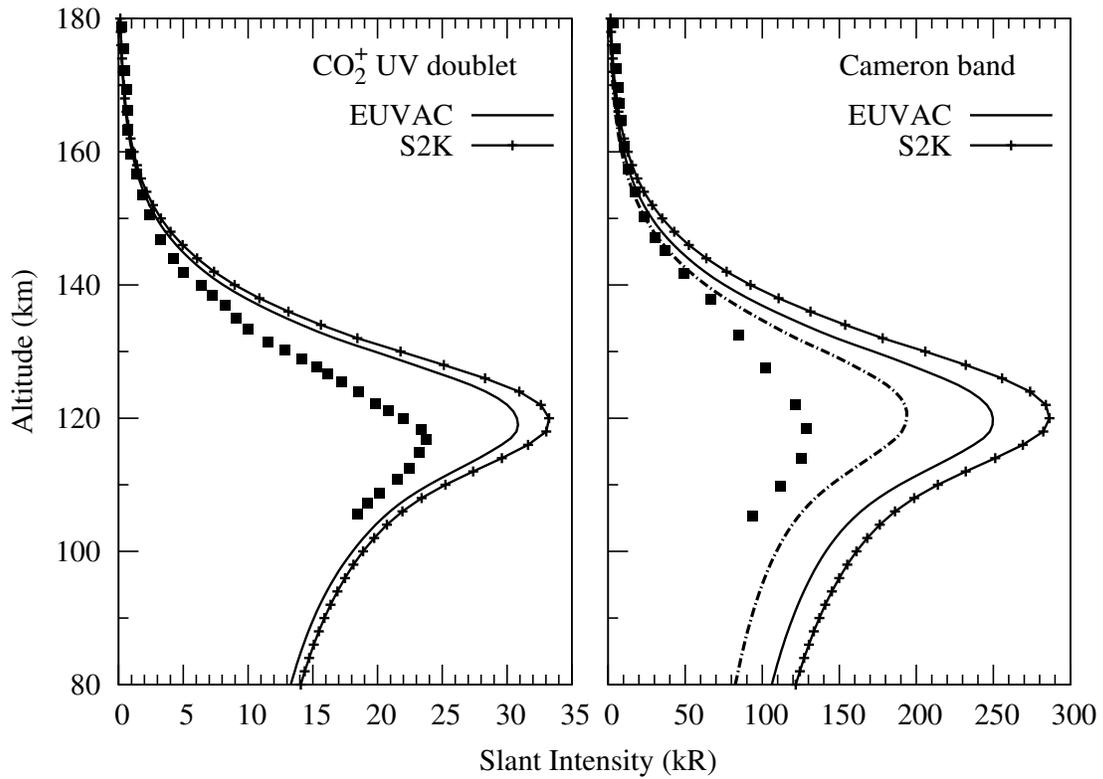}
\caption{Calculated limb profile of  \carp\ UV doublet band (left panel) and CO
Cameron band (right panel) for Ls $<$ 130\dgr. Symbols represent the SPICAM-observed
values taken from \cite{Leblanc06}. Dash-dotted curve shows the calculated intensity of
Cameron band with reduced (by a factor of 2) e-\car\ cross section.}
\label{fig:int-min}
\end{figure}

\begin{figure}
\centering
\includegraphics[width=36pc]{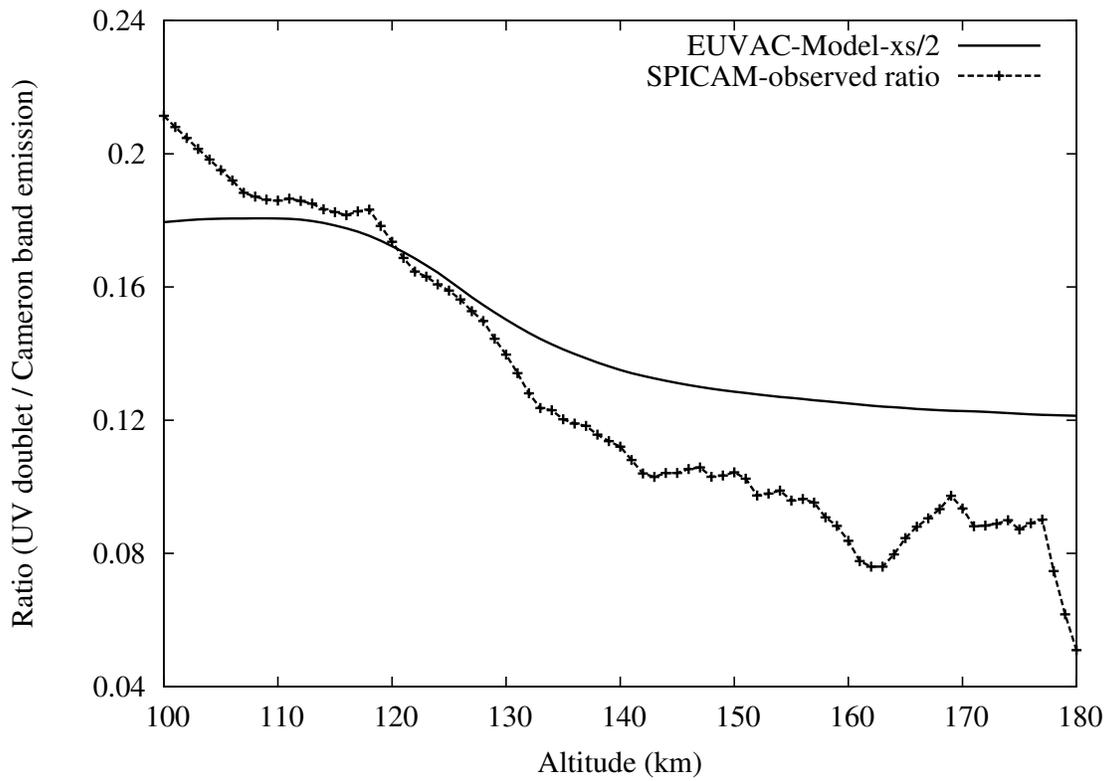}
\caption{Altitude variation of intensity ratio of UV doublet and CO Cameron band system.
Calculated
ratio is shown for EUVAC solar flux model with Cameron band production cross section in e-\car\ 
collision is reduced by a factor of 2. SPICAM-observed ratio is from Fig.~9(a) of \cite{Leblanc06}.}
\label{fig:ratio-min}
\end{figure}

\begin{figure}
\centering
\includegraphics[width=30pc]{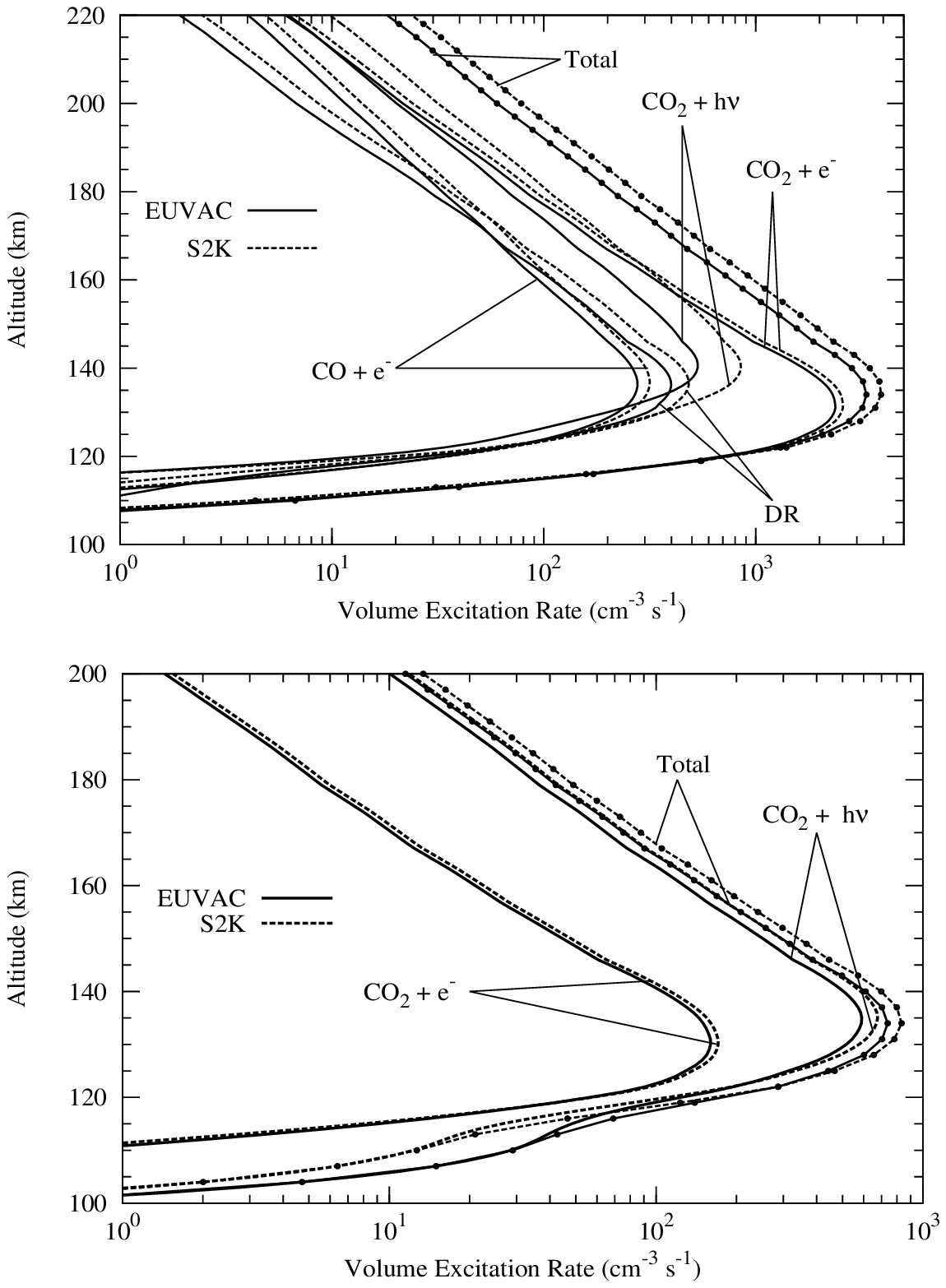}
\caption{Calculated production rates of the \coa\ (upper panel) and \carpb\ (bottom panel)
for solar longitude Ls $>$ 130\dgr. DR stands for dissociative recombination.}
\label{fig:ver-cam-138}
\end{figure}

\begin{figure}
\centering
\includegraphics[width=36pc]{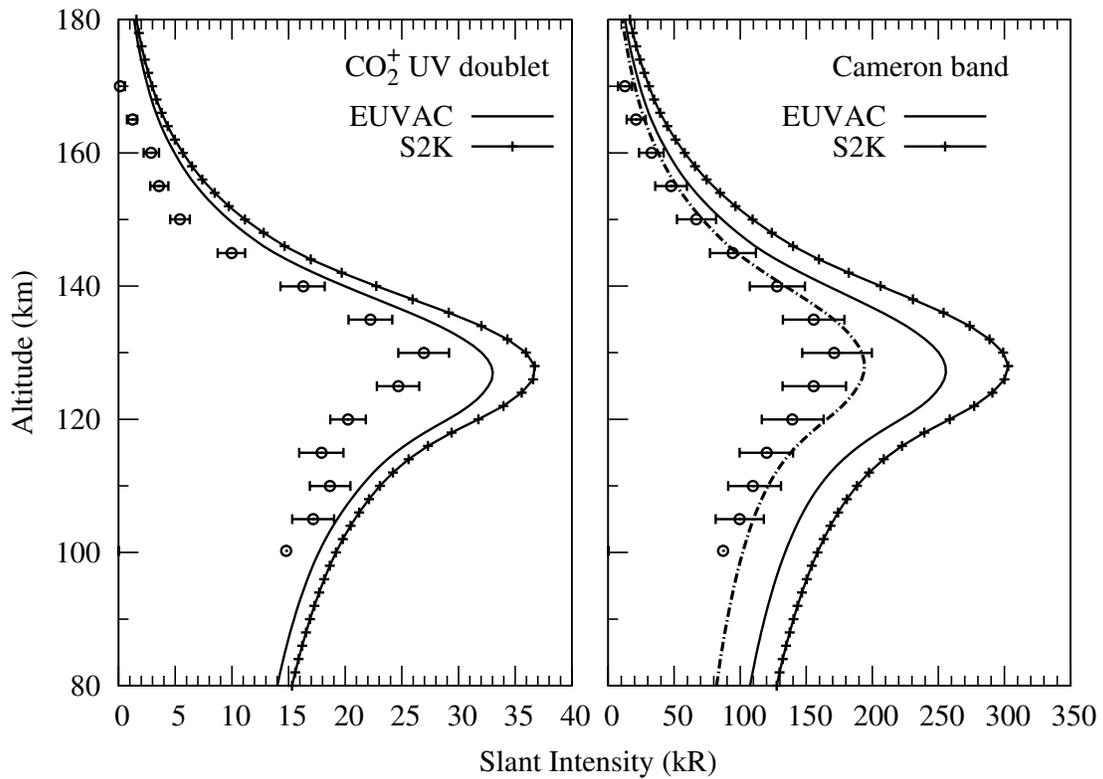}
\caption{Calculated limb profiles of \carp\ UV doublet band (left panel) and
CO Cameron band (right panel) for Ls $>$ 130\dgr.  Dash dotted curve shows the
Cameron band intensity for EUVAC model with e-\car\ cross section forming
\coa\ reduced by a factor of 2. Open circles with error bars represent
the SPICAM-observed intensity taken from \cite{Shematovich08}.}
\label{fig:int-max}
\end{figure}

\begin{figure}
\centering
\includegraphics[width=30pc]{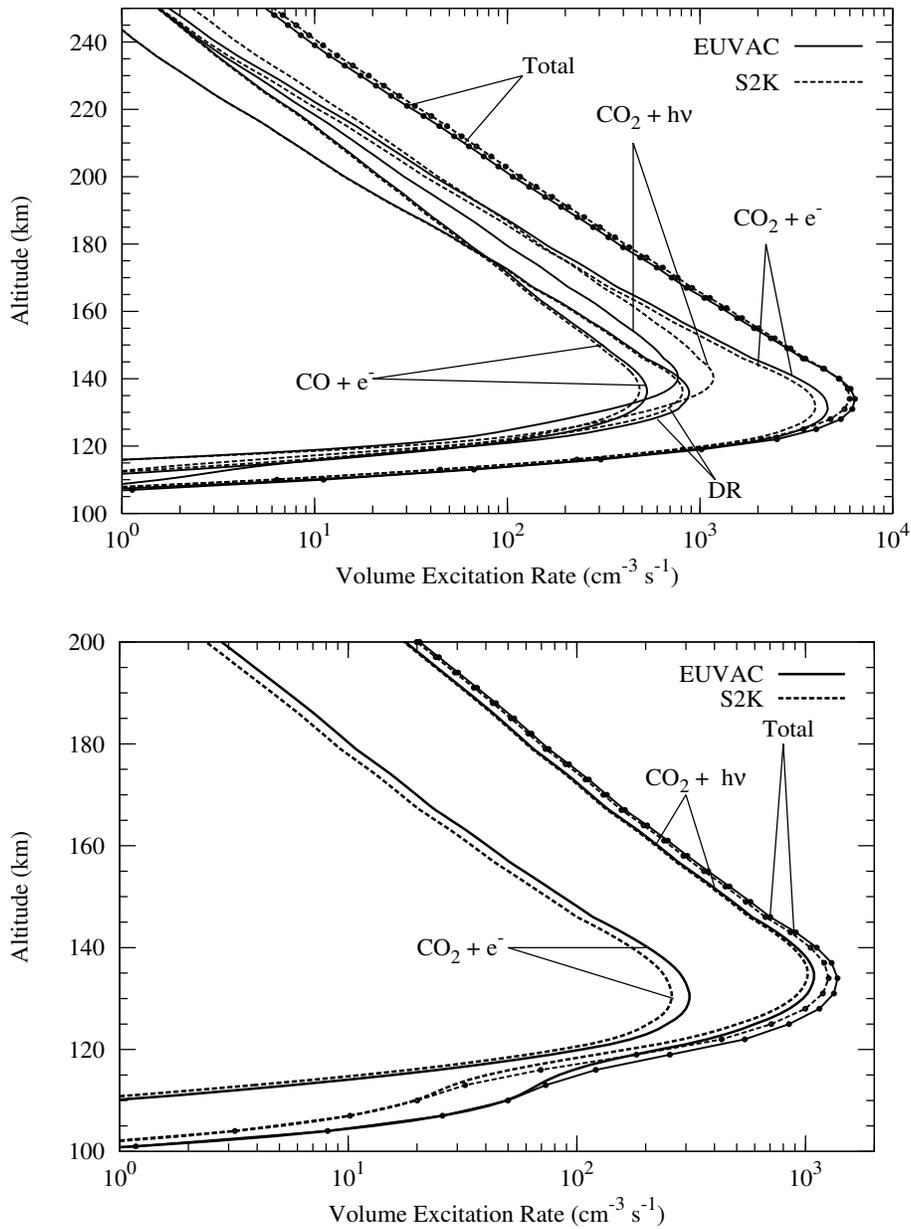}
\caption{Calculated production rates of the \coa\ (upper panel) and \carpb\ (bottom panel)
for solar maximum condition. DR stands for dissociative recombination.}
\label{fig:ver-cam-max}
\end{figure}

\begin{figure}
\centering
\includegraphics[width=36pc]{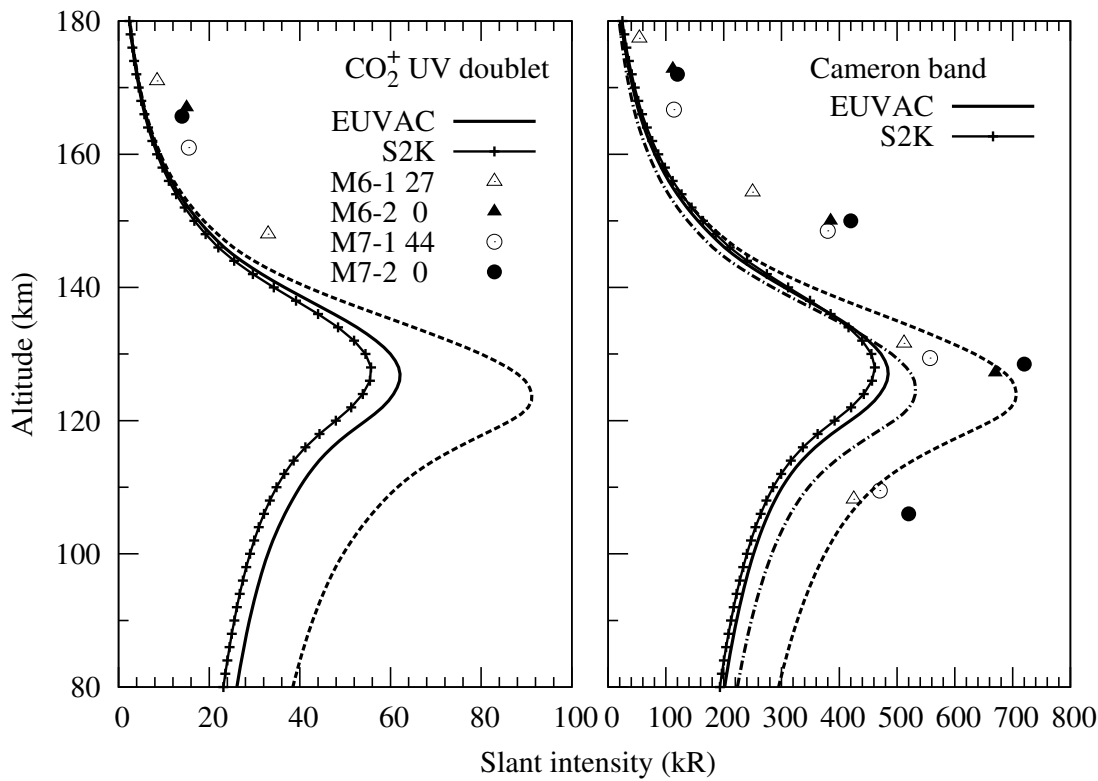}
\caption{Calculated limb intensity for the higher solar activity condition similar to 
Mariner 6 and 7 flybys. Solid curve shows the intensity calculated using EUVAC model at
SZA = 45\dgr. Solid curve with symbols shows the limb 
intensity calculated using S2K model at SZA = 45\dgr. Dashed curve shows the calculated
intensity (using EUVAC model) at SZA = 0\dgr. Dash dotted curve shows the calculated
intensity (using EUVAC model) at SZA = 0\dgr\ and after reducing the e-\car\ cross sections 
forming \coa\ by a factor of 2.  Symbol represents the observed intensity of
Cameron band and UV doublet band measured by Mariner 6 and 7 \citep{Stewart72a}. Observed values
are shown for 2 orbits each of Mariner 6 (for SZA = 27 and 0\dgr; open and solid triangle,
respectively) and Mariner 7 (for SZA = 44 and 0\dgr; open and solid circle, respectively).}
\label{fig:int-mariner}
\end{figure}

Calculations are also made for the high solar (Mariner 6 and 7 observations)
activity condition. Though the contribution of \car\ photodissociative excitation
in Cameron band production is higher when S2K model is used, but the total
intensity of CO Cameron band calculated using the EUVAC model is slightly higher
than that calculated using the S2K model. This is because of the higher
photoelectron flux when EUVAC model is used (Fig.~\ref{fig:pef}). 
The calculated intensities of both Cameron and UV doublet band are lower than the
observed values (Fig.~\ref{fig:int-mariner}).

Following conclusion can be drawn from the present study:
\begin{enumerate}
\item Generally, solar EUV fluxes in bands are higher in S2K model except at few
bands at shorter wavelength range ($<$ 250 \AA). Solar EUV fluxes at longer wavelengths
are higher in S2K model, specially in the 1000-1050 \AA\ bin, where flux is
around an order of magnitude higher than the corresponding flux in EUVAC model.
Solar EUV flux at lines is smaller in the S2K model compared to that in the 
EUVAC model.

\item Due to higher EUV flux at lines in the EUVAC model, peaks at 20--30 eV
range in the photoelectron flux are more prominent when EUVAC model is used.

\item During high solar activity condition, calculated photoelectron fluxes
are higher for EUVAC model due to higher EUV fluxes below 250~\AA\ in the EUVAC model.
Hence, intensities calculated using EUVAC model are higher by 5--10\% than
those calculated using S2K model.

\item During solar minimum condition, the Cameron and \carp\ UV doublet
intensities calculated using S2K solar flux model are $\sim$30--40\% higher
than those calculated using the EUVAC model.

\item During both, solar minimum as well as maximum conditions, Cameron band
production due to photodissociative excitation of \car\ 
is about 50\% higher when S2K solar EUV flux model is used.

\item Altitude of peak production rate of Cameron and \carp\ UV doublet bands
is independent of solar EUV flux model used in the calculations. However, for the 
Cameron band the altitude where photodissociation of \car\ takes over electron
impact dissociation is higher in the EUVAC model compared to that in the S2K model.

\item Reduction in the e-\car\ cross section producing Cameron
band and photoionization cross section producing \carp\ UV doublet band is
required to bring the model calculations in agreement with the observations.

\item For a given set of observation, and accounting for the uncertainties in
the cross sections, intensities calculated using the EUVAC model
are in better agreement with the observation than those calculated using the S2K model.
\end{enumerate}
Simultaneous observation of solar EUV flux with dayglow measurements
would be very helpful in improving our understanding of the processes that governs
the dayglow emissions on Mars. More accurate measurements of cross sections for electron
impact dissociation of \car\ producing Cameron band and photoionization of \car\ in $ B $
state are required for the better modelling of CO Cameron band and \carp\  UV doublet
band in Martian atmosphere, as well as in other \car--containing atmospheres, like Venus
and comets.

\end{multicols}

\newpage

\renewcommand{\thefootnote}{\fnsymbol{footnote}}

\begin{center}
\begin{table} 
\caption{Major reactions for the production and loss of CO($a^3\Pi$).}
\label{tab:rate-cameron}
\begin{tabular}{llp{1.5	in}}
\hline\noalign{\smallskip}
\multicolumn{1}{l}{ Reaction}
&\multicolumn{1}{l}{ Rate (cm$^{3}$ s$^{-1}$ or s$^{-1}$)}
&\multicolumn{1}{l}{ Reference} \\\noalign{\smallskip}
\hline\noalign{\smallskip}
CO$_2$ + h$\nu$ $\rightarrow$ CO(a$^3\Pi$) + O($^3$P)
&Model (K$_1$) & \cite{Lawrence72a} \\
CO$_2$ + e$^-_{ph}$ $\rightarrow$ CO(a$^3\Pi$) + O + e$^-$
& Model (K$_2$) &\textit{Present work}\\
CO + e$^-_{ph}$  $\rightarrow$ CO(a$^3\Pi$) + e$^-$ 
 &Model &\textit{Present work}\\
CO$_2^+$  + e$^-$ $\rightarrow$ CO(a$^3\Pi$) + O  
&K$_3$\footnotemark[2]
 &
\cite{Seiersen03, Rosati03}\\
CO(a$^3\Pi$) + CO$_2$ $\rightarrow$ CO + CO$_2$  &1.0 $\times$ 10$^{-11}$  
&\cite{Skrzypkowski98} \\
CO(a$^3\Pi$) + CO $\rightarrow$ CO + CO  &5.7 $\times$ 10$^{-11}$ 
&\cite{Wysong00}\\
 CO(a$^3\Pi$)\hspace{0.3cm}  $\longrightarrow$ CO  +  h$\nu$ & K$_4$= 1.26 $\times$ 10$^{2}$  
& \cite{Lawrence71} \\
\hline
\end{tabular}
\\
\small e$^-_{ph}$ = photoelectron.\\
\footnotemark[2]{\small K$_3 = $ 6.5 $\times$ 10$^{-7}$ (300/Te)$^{0.8}$ $\times$ 0.87 $\times$ 0.29 cm$^{3}$ s$^{-1}$; here 0.87 is yield of dissociative recombination 
of CO$_2^+$ producing CO, and 0.29 is yield of CO(a$^3\Pi$) produced from  CO.} \\ 
\small $K_1$, $K_2$, $K_3$, and $K_4$ are described in equation~\ref{eq:coa-den}.\\
\end{table}
\end{center}

\begin{center}
\begin{table}
\caption{Production rates (in cm$ ^{-3} $ s$ ^{-1} $) with peak altitude
for low solar activity condition (similar to Viking condition).}

\label{tab:ver}
\begin{tabular}{lll}
\hline \noalign{\smallskip}
Production Process & Cameron band & 
\multicolumn{1}{c}{\carp\ UV doublet}\\ \noalign{\smallskip}
\hline\noalign{\smallskip}
\car\ + h$ \nu $	& 465 at  137 km & 391 at 131 km \\
\car\ + e$^-_{ph}$	& 1470 at 128 km & 95  at 127 km \\
CO + e$^-_{ph}$		& 81 at 130 km   &	-			 \\
\carp\ + e$^-$		& 267 at 133 km	 &	-			 \\
Total				& 2094 at 131 km & 478 at 131 km \\
\hline
\end{tabular}
\\
\small e$^-_{ph}$ = photoelectron.
\end{table}
\end{center}

\begin{center}
	\begin{table}
	\footnotesize
		\caption{Overhead intensities (in kR) of CO Cameron and \carp\ UV doublet bands.}
		\smallskip
		\label{tab:oi-cmp}
		\begin{tabular}{lccccccccccc}
			\hline \noalign{\smallskip}
			\multirow{2}{1.8cm}{\centering Emissions} & 
			\multicolumn{5}{c}{EUVAC} & &
			\multicolumn{5}{c}{SOLAR2000} \\

			\cline{2-6}\cline{8-12} \noalign{\smallskip}
			& h$\nu$+\car\ & e+\car & DR\footnotemark[4] & 
			e+CO & Total &  &
			h$\nu$+\car & e+\car & DR & 
			e+CO & Total \\ 
			\hline  \noalign{\smallskip}
			& \multicolumn{11}{c}{Low solar activity (Viking condition)} \\ \noalign{\smallskip}
			Cameron band & 1.2 & 3.7 & 0.6 & 0.2 & 5.8  & &
			1.8 & 5 & 1  & 0.3 & 8.2 \\
			UV doublet & 1.1 & 0.2 & - & - & 1.3 & &
			1.4 & 0.3 & - & - & 1.7 \\ \noalign{\smallskip}

			& \multicolumn{11}{c}{SPICAM/Mars-Express, First Case (Ls $<$ 130\dgr)} \\ \noalign{\smallskip}
			Cameron band & 1.4 & 5.4 & 1.6 & 0.4 & 8.8 & &
			2.1 & 5.7 & 1.8 & 0.4 & 10.1 \\
			UV doublet & 1.4 & 0.4 & -   & -  & 1.8 & &
			1.6 & 0.4 & - & -  & 2 \\ \noalign{\smallskip}
			
			& \multicolumn{11}{c}{SPICAM/Mars-Express,  Second Case (Ls $>$ 130\dgr)} \\ \noalign{\smallskip}
			Cameron band & 1.6 & 6.1 & 1.1 & 0.9 & 9.8 &  &
			2.6 & 6.7 & 1.3 & 1 & 11.7 \\
			UV doublet & 1.7 & 0.4 & - & - &	2.1 & &
			1.9 & 0.4 & - & - & 2.3 \\ \noalign{\smallskip}
			
			& \multicolumn{11}{c}{Higher solar activity (Mariner observations)} \\ \noalign{\smallskip}
			Cameron band & 2.3 & 11.8 & 2.4 & 1.7 & 18.3 & &
			3.6 & 10.2 & 2.2 & 1.6 & 17.7 \\
			UV doublet & 3.1 & 0.8 & - & - & 3.9 & &
			2.8 & 0.7 & - & - & 3.5\\
			
			\hline
		\end{tabular}
		\\
		\footnotemark[4]{\small Dissociative recombination (e + \carp).} \\

	\end{table}
\end{center}

\begin{center}
\begin{table}
\caption{Overhead  and limb intensities of dif{}ferent vibrational
bands of CO Cameron band system calculated after reducing the e-\car\ cross section
by a factor of 2 in the first case (Ls $<$ 130\dgr).}
\label{tab:vib-oi}
\smallskip
\begin{tabular}{lccccccccc}
\hline \noalign{\smallskip}
Band & Band Origin & \multicolumn{2}{c}{Overhead (R)} & &
\multicolumn{2}{c}{Limb (kR)\footnotemark[2]} \\
\cline{3-4}\cline{6-7}
($ \nu' - \nu''$) & \AA\ & EUVAC & S2K &  & EUVAC & S2K \\
\hline \noalign{\smallskip}
0 - 0 & 2063.0 & 716 & 845 & & 20 & 24   \\
0 - 1 & 2158.4 & 713 & 840 & & 20 & 23.2 \\
0 - 2 & 2261.0 & 311 & 367 & & 9  & 10   \\
0 - 3 & 2374.0 &  79 & 93  & & 2  & 2.6  \\
1 - 0 & 1992.5 & 1151& 1358& & 32 & 38   \\
2 - 0 & 1927.5 & 603 & 711 & & 17 & 20   \\
3 - 0 & 1868.1 & 181 & 213 & & 5  & 6    \\
4 - 0 & 1813.0 & 38  & 45  & & 1  & 1.3  \\ \noalign{\smallskip}
\hline
\end{tabular}
\\ 
\footnotemark[2]{\small Limb intensities are given at altitude (120 km) of peak emission.}
\end{table}
\end{center}

\clearpage
\newpage

\begin{thebibliography}{66}
\expandafter\ifx\csname natexlab\endcsname\relax\def\natexlab#1{#1}\fi
\expandafter\ifx\csname url\endcsname\relax
  \def\url#1{\texttt{#1}}\fi
\expandafter\ifx\csname urlprefix\endcsname\relax\def\urlprefix{URL }\fi
  \providecommand{\doi}[1]{doi:\discretionary{}{}{}#1}

\bibitem[{Ajello(1971)}]{Ajello71b}
Ajello, J.~M., 1971. {Emission cross sections of CO$_2$ by electron impact in
  the interval 1260-4500 \AA}. J. Chem. Phys. 55, 3169 -- 3177.
\newblock \doi{\bibinfo{doi}{10.1063/1.1676564}}.

\bibitem[{Avakyan et~al.(1998)Avakyan, II'in, Lavrov, and Ogurtsov}]{Avakyan98}
Avakyan, S.~V., II'in, R.~N., Lavrov, V.~M., Ogurtsov, G.~N., 1998. In:
  Avakyan, S.~V. (Ed.), {Collision Processes and Excitation of UV Emission from
  Planetary Atmospheric Gases: A Handbook of Cross Sections}. Gordon and Breach
  science publishers.

\bibitem[{Barth et~al.(1971)Barth, Hord, Pearce, Kelly, Anderson, and
  Stewart}]{Barth71}
Barth, C.~A., Hord, C.~W., Pearce, J.~B., Kelly, K.~K., Anderson, G.~P.,
  Stewart, A.~I., 1971. {Mariner 6 and 7 ultraviolet spectrometer experiment:
  Upper atmosphere data}. J. Geophys. Res. 76, 2213 -- 2227.
\newblock \doi{\bibinfo{doi}{10.1029/JA076i010p02213}}.

\bibitem[{Bhardwaj(1999)}]{Bhardwaj99a}
Bhardwaj, A., 1999. {On the role of solar EUV, photoelectrons, and auroral
  electrons in the chemistry of C($^1$D) and the production of CI 1931 {\AA} in
  the inner cometary coma: A case for comet P/Halley}. J. Geophys. Res. 104,
  1929 -- 1942.
\newblock \doi{\bibinfo{doi}{10.1029/1998JE900004}}.

\bibitem[{Bhardwaj(2003)}]{Bhardwaj03}
Bhardwaj, A., 2003. {On the solar EUV deposition in the inner comae of comets
  with large gas production rates}. Geophys. Res. Lett. 30~(24), 2244.
\newblock \doi{\bibinfo{doi}{10.1029/2003GL018495}}.

\bibitem[{Bhardwaj et~al.(1990)Bhardwaj, Haider, and Singhal}]{Bhardwaj90b}
Bhardwaj, A., Haider, S.~A., Singhal, R.~P., 1990. {Auroral and photoelectron
  fluxes in cometary ionospheres}. Icarus 85, 216 -- 228.
\newblock \doi{\bibinfo{doi}{10.1016/0019-1035(90)90112-M}}.

\bibitem[{Bhardwaj et~al.(1996)Bhardwaj, Haider, and Singhal}]{Bhardwaj96}
Bhardwaj, A., Haider, S.~A., Singhal, R.~P., 1996. {Production and emissions of
  atomic carbon and oxygen in the inner coma of comet 1P/Halley: role of
  electron impact}. Icarus 120, 412 -- 430.
\newblock \doi{\bibinfo{doi}{10.1006/icar.1996.0061}}.

\bibitem[{Bhardwaj and Jain(2009)}]{Bhardwaj09}
Bhardwaj, A., Jain, S.~K., 2009. {Monte Carlo model of electron energy
  degradation in a CO$_2$ atmosphere}. J. Geophys. Res. 114.
\newblock \doi{\bibinfo{doi}{10.1029/2009JA014298}}.

\bibitem[{Bhardwaj and Jain(2011)}]{Bhardwaj11b}
Bhardwaj, A., Jain, S.~K., 2011. {Calculations of N$_2$ triplet states
  vibrational populations and band emissions in venusian dayglow}. Icarus .
\newblock \doi{\bibinfo{doi}{10.1016/j.icarus.2011.05.026}}.

\bibitem[{Bhardwaj and Michael(1999{\natexlab{a}})}]{Bhardwaj99d}
Bhardwaj, A., Michael, M., 1999{\natexlab{a}}. {Monte Carlo model for electron
  degradation in SO$_2$ gas: cross sections, yield spectra and efficiencies}.
  J. Geophys. Res. 104~(10), 24713 -- 24728.
\newblock \doi{\bibinfo{doi}{10.1029/1999JA900283}}.

\bibitem[{Bhardwaj and Michael(1999{\natexlab{b}})}]{Bhardwaj99b}
Bhardwaj, A., Michael, M., 1999{\natexlab{b}}. {On the excitation of Io's
  atmosphere by the photoelectrons: Application of the analytical yield
  spectrum of SO$_2$}. Geophys. Res. Lett. 26, 393 -- 396.
\newblock \doi{\bibinfo{doi}{10.1029/1998GL900320}}.

\bibitem[{{Bhardwaj} and {Raghuram}(2011)}]{Bhardwaj11a}
{Bhardwaj}, A., {Raghuram}, S., 2011. {Model for Cameron-band emission in
  comets: A case for the EPOXI mission target comet 103P/Hartley 2}. Mon. Not.
  R. Astron. Soc. Lett. 412, L25 -- L29.
\newblock \doi{\bibinfo{doi}{10.1111/j.1745-3933.2010.00998.x}}.

\bibitem[{Bhardwaj and Singhal(1993)}]{Bhardwaj93}
Bhardwaj, A., Singhal, R.~P., 1993. {Optically thin H Lyman alpha production on
  outer planets: Low-energy proton acceleration in parallel electric fields and
  neutral H atom precipitation from ring current}. J. Geophys. Res. 98~(A6),
  9473 -- 9481.
\newblock \doi{\bibinfo{doi}{10.1029/92JA02400}}.

\bibitem[{Bougher et~al.(2004)Bougher, Engel, Hinson, and Murphy}]{Bougher04}
Bougher, S.~W., Engel, S., Hinson, D.~P., Murphy, J.~R., 2004. {MGS Radio
  Science electron density profiles: Interannual variability and implications
  for the Martian neutral atmosphere}. J. Geophys. Res. 109.
\newblock \doi{\bibinfo{doi}{10.1029/2003JE002154}}.

\bibitem[{Bougher et~al.(1999)Bougher, Engel, Roble, and Foster}]{Bougher99}
Bougher, S.~W., Engel, S., Roble, R.~G., Foster, B., 1999. {Comparative
  terrestrial planet thermospheres: 2. Solar cycle variation of global
  structure and winds at equinox}. J. Geophys. Res. 104, 16591 -- 16611.
\newblock \doi{\bibinfo{doi}{10.1029/1998JE001019}}.

\bibitem[{Bougher et~al.(2000)Bougher, Engel, Roble, and Foster}]{Bougher00}
Bougher, S.~W., Engel, S., Roble, R.~G., Foster, B., 2000. {Comparative
  terrestrial planet thermospheres: 3. Solar cycle variation of global
  structure and winds at solstices}. J. Geophys. Res. 105, 17669 -- 17692.
\newblock \doi{\bibinfo{doi}{10.1029/1999JE001232}}.

\bibitem[{Buonsanto et~al.(1995)Buonsanto, Richards, Tobiska, Solomon, Tung,
  and Fennelly}]{Buonsanto95}
Buonsanto, M.~J., Richards, P.~G., Tobiska, W.~K., Solomon, S.~C., Tung, Y.
  .-K., Fennelly, J.~A., 1995. {Ionospheric Electron Densities Calculated Using
  Different EUV Flux Models and Cross Sections: Comparison with Radar Data}. J.
  Geophys. Res. 100~(A8), 14569 -- 14580.
\newblock \doi{\bibinfo{doi}{10.1029/95JA00680}}.

\bibitem[{Chaufray et~al.(2008)Chaufray, Bertaux, Leblanc, and
  Qu{\'e}merais}]{Chaufray08}
Chaufray, J.~Y., Bertaux, J.~L., Leblanc, F., Qu{\'e}merais, E., 2008.
  {Observation of the hydrogen corona with SPICAM on Mars Express}. Icarus
  195~(2), 598 -- 613.
\newblock \doi{\bibinfo{doi}{10.1016/j.icarus.2008.01.009}}.

\bibitem[{Conway(1981)}]{Conway81}
Conway, R.~R., 1981. {Spectroscopy of the Cameron bands in the Mars airglow}.
  J. Geophys. Res. 86, 4767 -- 4775.
\newblock \doi{\bibinfo{doi}{10.1029/JA086iA06p04767}}.

\bibitem[{Cox et~al.(2010)Cox, G{\'e}rard, Hubert, Bertaux, and
  Bougher}]{Cox10}
Cox, C., G{\'e}rard, J.~C., Hubert, B., Bertaux, J.~L., Bougher, S.~W., 2010.
  {Mars ultraviolet dayglow variability: SPICAM observations and comparison
  with airglow model}. J. Geophys. Res. 115.
\newblock \doi{\bibinfo{doi}{10.1029/2009JE003504}}.

\bibitem[{Erdman and Zipf(1983)}]{Erdman83}
Erdman, P.~W., Zipf, E.~C., 1983. {Electron-impact excitation of the Cameron
  system ($a^3\Pi\rightarrow X^1\Sigma$) of CO}. Planet. Spece. Sci. 31, 317 --
  321.
\newblock \doi{\bibinfo{doi}{10.1016/0032-0633(83)90082-X}}.

\bibitem[{Forget et~al.(2009)Forget, Montmessin, Bertaux, Galindo, Lebonnois,
  Qu{\'e}merais, Reberac, Dimarellis, and Valverde}]{Forget09}
Forget, F., Montmessin, F., Bertaux, J.~L., Galindo, F.~G., Lebonnois, S.,
  Qu{\'e}merais, E., Reberac, A., Dimarellis, E., Valverde, M. A.~L., 2009.
  {Density and temperatures of the upper Martian atmosphere measured by stellar
  occultations with Mars Express SPICAM}. J. Geophys. Res. 114.
\newblock \doi{\bibinfo{doi}{10.1029/2008JE003086}}.

\bibitem[{Fox and Sung(2001)}]{Fox01}
Fox, J., Sung, K., 2001. {Solar activity variations of the Venus
  thermosphere/ionosphere}. J. Geophys. Res. 106~(A10), 21305 -- 21335.
\newblock \doi{\bibinfo{doi}{10.1029/2001JA000069}}.

\bibitem[{Fox(2004)}]{Fox04}
Fox, J.~L., 2004. {Response of the Martian thermosphere/ionosphere to enhanced
  fluxes of solar soft X rays}. J. Geophys. Res. 109.
\newblock \doi{\bibinfo{doi}{10.1029/2004JA010380}}.

\bibitem[{Fox(2009)}]{Fox09}
Fox, J.~L., 2009. {Morphology of the dayside ionosphere of Mars: Implications
  for ion outflows}. J. Geophys. Res. 114.
\newblock \doi{\bibinfo{doi}{10.1029/2009JE003432}}.

\bibitem[{Fox and Dalgarno(1979)}]{Fox79}
Fox, J.~L., Dalgarno, A., 1979. {Ionization, luminosity, and heating of the
  upper atmosphere of Mars}. J. Geophys. Res. 84, 7315 -- 7333.
\newblock \doi{\bibinfo{doi}{10.1029/JA084iA12p07315}}.

\bibitem[{Fox et~al.(1996)Fox, Zhou, and Bougher}]{Fox96}
Fox, J.~L., Zhou, P., Bougher, S.~W., 1996. The martian thermosphere/ionosphere
  at high and low solar activities. Adv. Space Res. 17~(11), 203 -- 218.
\newblock \doi{\bibinfo{doi}{10.1016/0273-1177(95)00751-Y}}.

\bibitem[{Freund(1971)}]{Freund71}
Freund, R.~S., 1971. {Dissociation of CO$_2$ by electron impact with the
  formation of metastable CO($a^3\Pi$) and O($^5$S)}. J. Chem. Phys. 55, 3569
  -- 3577.
\newblock \doi{\bibinfo{doi}{10.1063/1.1676615}}.

\bibitem[{Gilijamse et~al.(2007)Gilijamse, Hoekstra, Meek, Mets\"al\"a, van~de
  Meerakker, T, Meijer, Groenenboom, and C.}]{Gilijamse07}
Gilijamse, J.~J., Hoekstra, S., Meek, S.~A., Mets\"al\"a, M., van~de Meerakker,
  S. Y.~T., T, S.~Y., Meijer, G., Groenenboom, G.~C., C., G., 2007. {The
  radiative lifetime of metastable CO (a$^3\Pi, \nu$=0)}. J. Chem. Phys. 127,
  221102--4.
\newblock \doi{\bibinfo{doi}{10.1063/1.2813888}}.

\bibitem[{Haider and Bhardwaj(2005)}]{Haider05}
Haider, S.~A., Bhardwaj, A., 2005. {Radial distribution of production rates,
  loss rates and densities corresponding to ion masses $\le$40 amu in the inner
  coma of Comet Halley: Composition and chemistry}. Icarus 177, 196 -- 216.
\newblock \doi{\bibinfo{doi}{10.1016/j.icarus.2005.02.019}}.

\bibitem[{Halmann et~al.(1966)Halmann, M, Laulicht, and I}]{Halmann66}
Halmann, M, Laulicht, I, February 1966. {Isotope effects on vibrational
  transition probabilities.IV. Electronic transitions of isotopic C$_2$, CO,
  CN, H$_2$, and CH molecules}. Astrophysical Journal Supplement 12, 307 --
  321.
\newblock \doi{\bibinfo{doi}{10.1086/190130}}.

\bibitem[{Hinteregger(1976)}]{Hinteregger76}
Hinteregger, E., 1976. {EUV fluxes in the solar spectrum below 2000 \AA}. J .
  Atmos. Terr. Phys. 38, 791 -- 806.
\newblock \doi{\bibinfo{doi}{10.1016/0021-9169(76)90020-9}}.

\bibitem[{Hinteregger et~al.(1981)Hinteregger, Fukui, and
  Gilson}]{Hinteregger81}
Hinteregger, H.~E., Fukui, K., Gilson, B.~R., 1981. {Observational, reference
  and model data on solar EUV, from measurements on AE-E}. Geophys. Res. Lett.
  8~(11), 1147 -- 1150.
\newblock \doi{\bibinfo{doi}{10.1029/GL008i011p01147}}.

\bibitem[{{Huestis} et~al.(2010){Huestis}, {Slanger}, {Sharpee}, and
  {Fox}}]{Huestis10}
{Huestis}, D.~L., {Slanger}, T.~G., {Sharpee}, B.~D., {Fox}, J.~L., 2010.
  {Chemical origins of the Mars ultraviolet dayglow}. Faraday Discussions 147,
  307 -- 322.
\newblock \doi{\bibinfo{doi}{10.1039/c003456h}}.

\bibitem[{Itikawa(2002)}]{Itikawa02}
Itikawa, Y., 2002. {Cross sections for electron collisions with carbon
  dioxide}. J. Phys. Chem. Ref. Data 31~(3), 749 -- 767.
\newblock \doi{\bibinfo{doi}{10.1063/1.1481879}}.

\bibitem[{Jackman et~al.(1977)Jackman, Garvey, and Green}]{Jackman77}
Jackman, C., Garvey, R., Green, A., 1977. {Electron impact on atmospheric
  gases, I. Updated cross sections}. J. Geophys. Res. 82~(32), 5081 -- 5090.
\newblock \doi{\bibinfo{doi}{10.1029/JA082i032p05081}}.

\bibitem[{Jain and Bhardwaj(2011)}]{Jain11}
Jain, S.~K., Bhardwaj, A., 2011. {Model calculation of N$_2$ Vegard-Kaplan band
  emissions in Martian dayglow}. J. Geophys.
  Res.\doi{\bibinfo{doi}{10.1029/2010JE003778}}.

\bibitem[{Johnson(1972)}]{Johnson72}
Johnson, C.~E., 1972. {Lifetime of CO(a$^{3}\Pi$) following electron impact
  dissociation of CO$_2$}. J. Chem. Phys. 57~(1), 576 -- 577.
\newblock \doi{\bibinfo{doi}{10.1063/1.1678007}}.

\bibitem[{{Jongma} et~al.(1997){Jongma}, {Berden}, and {Meijer}}]{Jongma97}
{Jongma}, R.~T., {Berden}, G., {Meijer}, G., Nov. 1997. {State-specific
  lifetime determination of the a$^{3}\Pi$ state in CO}. J. Chem. Phys. 107,
  7034 -- 7040.
\newblock \doi{\bibinfo{doi}{10.1063/1.474946}}.

\bibitem[{Lawrence(1972)}]{Lawrence72a}
Lawrence, G., 1972. {Photodissociation of CO$_2$ to produce CO(a$^3\Pi$)}. J.
  Chem. Phys. 56, 3435 -- 3442.
\newblock \doi{\bibinfo{doi}{10.1063/1.1677717}}.

\bibitem[{{Lawrence}(1971)}]{Lawrence71}
{Lawrence}, G.~M., Jun. 1971. {Quenching and radiation rates of CO
  (a$^{3}\Pi$)}. Chem. Phys. Lett. 9, 575 -- 577.
\newblock \doi{\bibinfo{doi}{10.1016/0009-2614(71)85130-8}}.

\bibitem[{Lean(1990)}]{Lean90}
Lean, J., 1990. {A comparison of models of the Sun?s extreme ultraviolet
  irradiance variations}. J. Geophys. Res. 95~(A8), 11933 -- 11944.
\newblock \doi{\bibinfo{doi}{10.1029/JA095iA08p11933}}.

\bibitem[{Lean et~al.(2011)Lean, Woods, Eparvier, Meier, Strickland, Correira,
  and Evans}]{Lean11}
Lean, J.~L., Woods, T.~N., Eparvier, F.~G., Meier, R.~R., Strickland, D.~J.,
  Correira, J.~T., Evans, J.~S., 2011. {Solar extreme ultraviolet irradiance:
  Present, past, and future}. J. Geophys. Res. 116.
\newblock \doi{\bibinfo{doi}{10.1029/2010JA015901}}.

\bibitem[{Leblanc et~al.(2007)Leblanc, Chaufray, and Bertaux}]{Leblanc07}
Leblanc, F., Chaufray, J.~Y., Bertaux, J.~L., 2007. {On Martian nitrogen
  dayglow emission observed by SPICAM UV spectrograph/Mars Express}. Geophys.
  Res. Lett. 34.
\newblock \doi{\bibinfo{doi}{10.1029/2006GL0284}}.

\bibitem[{Leblanc et~al.(2006)Leblanc, Chaufray, Lilensten, Witasse, and
  Bertaux}]{Leblanc06}
Leblanc, F., Chaufray, J.~Y., Lilensten, J., Witasse, O., Bertaux, J.-L., 2006.
  {Martian dayglow as seen by the SPICAM UV spectrograph on Mars Express}. J.
  Geophys. Res. 111.
\newblock \doi{\bibinfo{doi}{10.1029/2005JE002664}}.

\bibitem[{Mantas and Hanson(1979)}]{Mantas79}
Mantas, G.~P., Hanson, W.~B., 1979. {Photoelectron fluxes in the Martian
  ionosphere}. J. Geophys. Res. 84, 369 -- 385.
\newblock \doi{\bibinfo{doi}{10.1029/JA084iA02p00369}}.

\bibitem[{Michael and Bhardwaj(1997)}]{Michael97}
Michael, M., Bhardwaj, A., 1997. {On the dissociative ionization of SO$_2$ in
  the Io's atmosphere}. Geophys. Res. Lett. 24, 1971 -- 1974.
\newblock \doi{\bibinfo{doi}{10.1029/97GL02056}}.

\bibitem[{Nier and McElroy(1976)}]{Nier76}
Nier, A.~O., McElroy, M.~B., 1976. {Structure of the neutral upper atmosphere
  of Mars: Results from Viking 1 and Viking 2}. Science 194, 1298 -- 1300.
\newblock \doi{\bibinfo{doi}{10.1126/science.194.4271.1298}}.

\bibitem[{Richards et~al.(1994)Richards, Fennelly, and Torr}]{Richards94}
Richards, P.~G., Fennelly, J.~A., Torr, D.~G., 1994. {EUVAC: A solar EUV flux
  model for aeronomic calculations}. J. Geophys. Res. 99, 8981 -- 8992.
\newblock \doi{\bibinfo{doi}{10.1029/94JA00518}}.

\bibitem[{Rosati et~al.(2003)Rosati, Johnsen, and Golde}]{Rosati03}
Rosati, R.~E., Johnsen, R., Golde, M.~F., 2003. {Absolute yields of CO ($a'^3
  \Sigma ^+$, $d^3 \Delta_i$, $e^3 \Sigma^-$) + O from the dissociative
  recombination of CO$_2^+$ ions with electrons}. J. Chem. Phys. 119, 11630 --
  11635.
\newblock \doi{\bibinfo{doi}{10.1063/1.1623480}}.

\bibitem[{Schunk and Nagy(2000)}]{Schunk00}
Schunk, R.~W., Nagy, A.~F., 2000. {Ionospheres: Physics, Plasma Physics, and
  Chemistry}. Cambridge University Press.

\bibitem[{Seiersen et~al.(2003)Seiersen, Al-Khalili, Heber, Jensen, Nielsen,
  Pedersen, Safvan, and Andersen}]{Seiersen03}
Seiersen, K., Al-Khalili, A., Heber, O., Jensen, M.~J., Nielsen, I.~B.,
  Pedersen, H.~B., Safvan, C.~P., Andersen, L.~H., Aug 2003. {Dissociative
  recombination of the cation and dication of CO$_2$}. Phys. Rev. A 68~(2),
  022708.
\newblock \doi{\bibinfo{doi}{10.1103/PhysRevA.68.022708}}.

\bibitem[{Shematovich et~al.(2008)Shematovich, Bisikalo, G{\'e}rard, Cox,
  Bougher, and Leblanc}]{Shematovich08}
Shematovich, V.~I., Bisikalo, D.~V., G{\'e}rard, J.-C., Cox, C., Bougher,
  S.~W., Leblanc, F., 2008. {Monte Carlo model of electron transport for the
  calculation of Mars dayglow emissions}. J. Geophys. Res. 113.
\newblock \doi{\bibinfo{doi}{10.1029/2007JE002938}}.

\bibitem[{Simon et~al.(2009)Simon, Witasse, Leblanc, Gronoff, and
  Bertaux}]{Simon09}
Simon, C., Witasse, O., Leblanc, F., Gronoff, G., Bertaux, J.-L., 2009.
  {Dayglow on Mars: Kinetic modeling with SPICAM UV limb data}. Planetary Space
  Sci. 57, 1008 -- 1021.
\newblock \doi{\bibinfo{doi}{10.1016/j.pss.2008.08.012}}.

\bibitem[{Singhal and Bhardwaj(1991)}]{Singhal91}
Singhal, R.~P., Bhardwaj, A., 1991. {Monte Carlo simulation of photoelectron
  energization in parallel electric fields: Electroglow on Uranus}. J. Geophys.
  Res. 96, 15963 -- 15972.
\newblock \doi{\bibinfo{doi}{10.1029/90JA02749}}.

\bibitem[{{Skrzypkowski} et~al.(1998){Skrzypkowski}, {Gougousi}, {Johnsen}, and
  {Golde}}]{Skrzypkowski98}
{Skrzypkowski}, M.~P., {Gougousi}, T., {Johnsen}, R., {Golde}, M.~F., May 1998.
  {Measurement of the absolute yield of CO(a$^{3}${$\Pi$})+O products in the
  dissociative recombination of CO$_2^{+}$ ions with electrons}. J. Chem. Phys.
  108, 8400 -- 8407.
\newblock \doi{\bibinfo{doi}{10.1063/1.476267}}.

\bibitem[{Stewart(1972)}]{Stewart72a}
Stewart, A.~I., 1972. {Mariner 6 and 7 ultraviolet spectrometer experiment:
  Implication of CO$_2^+$, CO, and O airglow}. J. Geophys. Res. 77, 54 -- 68.
\newblock \doi{\bibinfo{doi}{10.1029/JA077i001p00054}}.

\bibitem[{Stewart et~al.(1972)Stewart, Barth, Hord, and Lane}]{Stewart72b}
Stewart, A.~I., Barth, C.~A., Hord, C.~W., Lane, A.~L., 1972. {Mariner 9
  ultraviolet spectrometer experiment: Structure of Mars' upper atmosphere}.
  Icarus 17~(2), 469 -- 474.
\newblock \doi{\bibinfo{doi}{10.1016/0019-1035(72)90012-7}}.

\bibitem[{Tobiska and Barth(1990)}]{Tobiska90}
Tobiska, W., Barth, C., 1990. {A Solar EUV Flux Model}. J. Geophys. Res.
  95~(A6), 8243 -- 8251.
\newblock \doi{\bibinfo{doi}{10.1029/JA095iA06p08243}}.

\bibitem[{Tobiska(1991)}]{Tobiska91}
Tobiska, W.~K., 1991. {Revised Solar Extreme Ultraviolet Flux Model}. J. Atm.
  Terr. Phys. 53, 1005 -- 1018.
\newblock \doi{\bibinfo{doi}{10.1016/0021-9169(91)90046-A}}.

\bibitem[{Tobiska(1994)}]{Tobiska94}
Tobiska, W.~K., 1994. {Modeled soft X-ray solar irradiances}. Solar Phys. 152,
  207 -- 215.
\newblock \doi{\bibinfo{doi}{10.1007/BF01473206}}.

\bibitem[{Tobiska(2004)}]{Tobiska04}
Tobiska, W.~K., 2004. {SOLAR2000 irradiances for climate change, aeronomy and
  space system engineering}. Adv. Space Res. 34, 1736 -- 1746.
\newblock \doi{\bibinfo{doi}{10.1016/j.asr.2003.06.032}}.

\bibitem[{Tobiska et~al.(2000)Tobiska, Woods, Eparvier, Viereck, Floyd, Bouwer,
  Rottman, and White}]{Tobiska00}
Tobiska, W.~K., Woods, T., Eparvier, F., Viereck, R., Floyd, L., Bouwer, D.,
  Rottman, G., White, O.~R., 2000. {The SOLAR2000 empirical solar irradiance
  model and forecast tool}. J. Atmos. Sol. Terr. Phys. 62, 1233 -- 1250.
\newblock \doi{\bibinfo{doi}{10.1016/S1364-6826(00)00070-5}}.

\bibitem[{Torr and Torr(1985)}]{Torr85}
Torr, M.~R., Torr, D.~G., 1985. {Ionization frequencies for solar cycle 21 -
  revised}. J. Geophys.Res. 90, 6675 -- 6678.
\newblock \doi{\bibinfo{doi}{10.1029/JA090iA07p06675}}.

\bibitem[{Torr et~al.(1979)Torr, Torr, and Hinteregger}]{Torr79}
Torr, M.~R., Torr, D. G.~T., Hinteregger, H.~E., 1979. {Ionization freequencies
  for major thermospheric constituents as a function of solar cycle 21}.
  Geophys. Res. Lett. 6, 771 -- 774.
\newblock \doi{\bibinfo{doi}{10.1029/GL006i010p00771}}.

\bibitem[{Wysong(2000)}]{Wysong00}
Wysong, I.~J., 2000. {Measurement of quenching rates of CO(a$^3\Pi, \nu=0$)
  using laser pump-and-probe technique}. Chem. Phys. Lett. 329~(1-2), 42 -- 46.
\newblock \doi{\bibinfo{doi}{10.1016/S0009-2614(00)00967-2}}.

\end{thebibliography}
\end{document}